\documentclass[11pt,a4paper]{article}
\usepackage[utf8]{inputenc}
\usepackage{a4wide}
\usepackage{amsmath}
\usepackage{amssymb}
\usepackage{amsfonts}
\usepackage{cite}
\usepackage{color}
\usepackage{adjustbox}
\usepackage{multirow}
\usepackage{pdflscape}
\usepackage{textcomp}
\usepackage{gensymb}
\usepackage{graphicx}
\usepackage{diagbox}
\usepackage{pifont}
\usepackage{bigstrut}
\usepackage[small,bf]{caption}
\usepackage{tabulary}
\usepackage{booktabs}
\usepackage{stackengine}
\usepackage{mathtools}
\usepackage{enumerate}
\usepackage{makecell}
\usepackage{listings}
\usepackage{bbm}
\usepackage[unicode]{hyperref}
\newcommand{\be}{\begin{equation}}
\newcommand{\ee}{\end{equation}}
\def\1{\mathbf{1}}

\def\2{\mathbf{2}}
\def\3{\mathbf{3}}

\def\5{\mathbf{5}}

\def\g{\gamma}
\def\G{\Gamma}

\def\t{\tau}
\def\th{\theta}

\def\ot{\otimes}

\def\di{{\rm d}}

\DeclareMathOperator{\diag}{diag}

\DeclareMathOperator{\im}{Im}
\DeclareMathOperator{\re}{Re}

\allowdisplaybreaks
\numberwithin{equation}{section}
\makeatletter
\g@addto@macro\bfseries{\boldmath}
\makeatother
\begin{document}
%%%%%%%%%%%%%%%%%%%%%%%

%%%%%%%%%%%%%%%%%%%%%%%
\begin{titlepage}

\vspace*{-15mm}
\begin{flushright}
SISSA 14/2019/FISI \\
IPMU19-0080 \\
IPPP/19/46 \\
CFTP/19-019
\end{flushright}
\vspace*{8mm}

\begin{center}
{\bf\LARGE {Generalised CP Symmetry in\\[2mm]
Modular-Invariant Models of Flavour}}\\[8mm]
P.~P.~Novichkov$^{\,a,}$\footnote{E-mail: \texttt{pavel.novichkov@sissa.it}}, 
J.~T.~Penedo$^{\,b,}$\footnote{E-mail: \texttt{joao.t.n.penedo@tecnico.ulisboa.pt}}, 
S.~T.~Petcov$^{\,a,c,}$\footnote{Also at
Institute of Nuclear Research and Nuclear Energy,
Bulgarian Academy of Sciences, 1784 Sofia, Bulgaria.},
A.~V.~Titov$^{\,d,}$\footnote{E-mail: \texttt{arsenii.titov@durham.ac.uk}}\\
\vspace{8mm}
$^{a}$\,{\it SISSA/INFN, Via Bonomea 265, 34136 Trieste, Italy} \\
\vspace{2mm}
$^{b}$\,{\it CFTP, Departamento de Física, Instituto Superior Técnico, Universidade de Lisboa,\\
Avenida Rovisco Pais 1, 1049-001 Lisboa, Portugal} \\
\vspace{2mm}
$^{c}$\,{\it Kavli IPMU (WPI), University of Tokyo, 5-1-5 Kashiwanoha, 277-8583 Kashiwa, Japan} \\
\vspace{2mm}
$^{d}$\,{\it Institute for Particle Physics Phenomenology, 
Department of Physics, Durham University,\\ 
South Road, Durham DH1 3LE, United Kingdom}
\end{center}
\vspace{8mm}

\begin{abstract}
\noindent
The formalism of combined finite modular and 
generalised CP (gCP) symmetries for theories 
of flavour is developed.
The corresponding consistency conditions for the two symmetry transformations 
acting on the modulus $\tau$ and on the matter fields are derived.
The implications of gCP symmetry 
in theories of flavour based on modular invariance 
described by finite modular groups are illustrated 
with the example of a modular $S_4$ model of lepton flavour.
Due to the addition of the gCP symmetry, 
viable modular models turn out to be more constrained,
with the modulus $\tau$ being the only source of CP violation.
\end{abstract}

\end{titlepage}
\setcounter{footnote}{0}
%%%%%%%%%%%%%%%%%%%%%%%

%%%%%%%%%%%%%%%%%%%%%%%
\section{Introduction}
\label{sec:intro}
%%%%%%%%%%%%%%%%%%%%%%%
%
Explaining the flavour structures of quarks and leptons 
remains to be one of the fundamental problems in particle physics.
We still do not know the absolute values of neutrino masses, 
as well as whether CP symmetry is violated in the lepton sector. 
However, the observed pattern of neutrino mixing with two large and one small 
(but non-zero) angles suggests that a non-Abelian discrete flavour symmetry 
can be at work (see \cite{Altarelli:2010gt,Ishimori:2010au,King:2013eh,Petcov:2017ggy} for reviews). 

In the bottom-up discrete symmetry approach to lepton flavour, 
some of the neutrino mixing angles and the Dirac CP violation (CPV) phase $\delta$ 
are generically predicted to be correlated with each other, 
since all of them are expressed in terms of few free parameters. 
At the same time, the Majorana phases~\cite{Bilenky:1980cx}, 
present in the neutrino mixing matrix if neutrinos 
are Majorana particles, remain unconstrained. 
In order to reduce the number of free parameters, 
the discrete flavour symmetry can be combined with 
the so-called generalised CP (gCP) symmetry~\cite{Feruglio:2012cw,Holthausen:2012dk}. 
Such models have more predictive power and allow, in particular, 
for prediction of the Majorana phases. 
The implications of combining the gCP symmetry 
with a flavour symmetry have been extensively studied for many discrete groups, 
including $A_4$~\cite{Holthausen:2012dk,Ding:2013bpa}, 
$T'$~\cite{Girardi:2013sza},
$S_4$~\cite{Feruglio:2012cw,Ding:2013hpa,Feruglio:2013hia,Li:2013jya,Li:2014eia,Lu:2016jit,Penedo:2017vtf} 
and $A_5$~\cite{Li:2015jxa,DiIura:2015kfa,Ballett:2015wia,Turner:2015uta} (see also~\cite{Girardi:2016zwz}).

The conventional bottom-up discrete symmetry approach 
to lepton flavour has certain drawbacks. 
Within this approach specific models need to be constructed 
to obtain predictions for neutrino masses. 
A flavour symmetry in these models is typically spontaneously broken 
by vacuum expectation values (VEVs) of scalar flavon fields. 
Usually, a relatively large
number of these fields with a rather complicated potential 
possessing additional shaping symmetries
to achieve the correct vacuum alignment is needed. 
Possible higher-dimensional operators may affect model predictions, 
and thus, have to be taken into account. 

In view of that, a new approach, in which modular invariance plays the role 
of flavour symmetry, has been put forward in Ref.~\cite{Feruglio:2017spp}. 
The main feature of this approach is that the Yukawa couplings 
and fermion mass matrices
in the Lagrangian of the theory arise from
modular forms which depend on the value of a single 
complex scalar field $\tau$, called the modulus. 
In addition, both the couplings and fields transform under 
a finite modular group $\Gamma_N$.
Once $\tau$ acquires a VEV, the 
Yukawa couplings and the form of the mass matrices
get fixed, and a certain flavour structure arises. 
For $N \leq 5$, the finite modular groups are isomorphic to permutation groups 
(see, e.g., \cite{deAdelhartToorop:2011re}) used to 
build models of lepton and quark flavours. 
Until now models based on the finite modular groups 
$\Gamma_2 \simeq S_3$~\cite{Kobayashi:2018vbk,Kobayashi:2018wkl}, 
$\Gamma_3 \simeq A_4$~\cite{Feruglio:2017spp,Kobayashi:2018vbk,Criado:2018thu,Kobayashi:2018scp,deAnda:2018ecu,Okada:2018yrn,Kobayashi:2018wkl,Novichkov:2018yse}, 
$\Gamma_4 \simeq S_4$~\cite{Penedo:2018nmg,Novichkov:2018ovf} 
and $\Gamma_5 \simeq A_5$~\cite{Novichkov:2018nkm,Ding:2019xna} 
have been constructed in the literature. 
In a top-down approach,
the interplay of flavour and modular symmetries
has recently been considered
in the context of string theory
in Refs.~\cite{Kobayashi:2018rad,Kobayashi:2018bff,Baur:2019kwi}.

In the present work, we study the implications of combining 
the gCP symmetry with modular invariance
in the construction of models of flavour.
It is expected that combining the two symmetries 
in a model of flavour will lead to a reduction of the 
number of free parameters, and thus to an increased predictive power 
of the model.
The article is organised as follows. 
In Section~\ref{sec:framework}, we summarise 
key features of combining gCP symmetry 
with a discrete non-Abelian group 
and briefly describe the modular symmetry approach to flavour. 
Then, in Section~\ref{sec:modularGCP}, requiring consistency 
between CP and modular symmetries, 
we derive the action of CP on 
i)~the modulus, ii)~superfields and iii)~multiplets of modular forms. 
 After that, in Section~\ref{sec:CP_inv}, we discuss implications for the charged lepton and neutrino mass matrices,
and determine the values of the modulus which allow for CP conservation. 
In Section~\ref{sec:s4}, we give an example of a viable model 
invariant under both the modular and CP symmetries. 
Finally, we conclude in Section~\ref{sec:conclusions}.

%%%%%%%%%%%%%%%%%%%%%%%
\section{Framework}
\label{sec:framework}
%%%%%%%%%%%%%%%%%%%%%%%

%=======================
\subsection{Generalised CP Symmetry Combined with a Flavour Symmetry}
\label{sec:gcp_general}
%=======================

Consider a supersymmetric (SUSY) theory%
\footnote{In what follows we will focus on a supersymmetric construction 
with global supersymmetry (see subsection~\ref{sec:modular_sym}).}
with a flavour symmetry described by a 
non-Abelian discrete group $G_f$. 
A chiral superfield $\psi(x)$ in a generic irreducible 
representation (irrep) $\mathbf{r}$ of $G_f$ 
transforms under the action of $G_f$ as 
%%%%%%%
\be
\psi(x) \,\xrightarrow{g}\, \rho_\mathbf{r}(g)\, \psi(x)\,, \quad\, g \in G_f\,, 
\label{eq:flavourtransformation}
\ee
%%%%%%%
%
where $\rho_\mathbf{r}(g)$ is the unitary representation matrix 
for the element $g$ in the irrep $\mathbf{r}$. 
A theory which is also invariant under CP symmetry 
has to remain unchanged under the following transformation: 
%%%%%%%
\be
\psi(x) \,\xrightarrow{CP}\, X_\mathbf{r}\, \overline\psi(x_P)\,, 
\label{eq:GCPtransformation}
\ee
%%%%%%%
%
with a bar denoting the Hermitian conjugate superfield, and
where $x = (t,\mathbf{x})$, $x_P = (t,-\mathbf{x})$ and $X_\mathbf{r}$ is 
a unitary matrix acting on flavour space~\cite{Branco:1986gr}. 
The transformation in eq.~\eqref{eq:GCPtransformation} is commonly 
referred to as a gCP transformation.
In the case of $X_\mathbf{r} = \mathbbm{1}_\mathbf{r}$, 
one recovers the canonical CP transformation. 
The action of the gCP transformation on a chiral superfield 
and, in particular, on its fermionic component is described in detail 
in Appendix~\ref{app:SUSYCP}.

The form of the matrix $X_\mathbf{r}$ is constrained due to the presence of a 
flavour symmetry~\cite{Feruglio:2012cw,Holthausen:2012dk}. 
Performing first a gCP transformation, 
followed by a flavour symmetry transformation $g \in G_f$, and subsequently
an inverse gCP transformation, one finds
%%%%%%%
\be
\psi(x) \,\xrightarrow{CP}\, X_\mathbf{r}\, \overline\psi(x_P) \,\xrightarrow{g}\, 
X_\mathbf{r}\, \rho_\mathbf{r}^*(g) \overline\psi(x_P) \,\xrightarrow{CP^{-1}}\, 
X_\mathbf{r}\, \rho_\mathbf{r}^*(g) X_\mathbf{r}^{-1} \psi(x)\,.
\ee
%%%%%%%
%
The theory should remain invariant under this sequence of transformations, 
and thus, the resulting transformation must correspond to a flavour symmetry 
transformation (cf. eq.~\eqref{eq:flavourtransformation}) $\rho_\mathbf{r}(g')$, 
with $g'$ being some element of $G_f$, i.e., we have:
%%%%%
\be
X_\mathbf{r}\, \rho_\mathbf{r}^*(g) X_\mathbf{r}^{-1} = \rho_\mathbf{r}(g')\,, \quad g,\, g' \in G_f\,.
\label{eq:consistency}
\ee
%%%%%
%
This equation defines the consistency condition, which has to be respected for 
consistent implementation of a gCP symmetry along with a flavour symmetry,
provided the full flavour symmetry group $G_f$ has been correctly identified
\cite{Feruglio:2012cw,Holthausen:2012dk}. 
Notice that $X_\mathbf{r}$ is a unitary matrix defined for 
each irrep~\cite{Chen:2014tpa}.
Several well-known facts about this consistency condition are in order. 
%%%%%%%
\begin{itemize}
\item Equation~\eqref{eq:consistency} has to be satisfied for all 
irreps $\mathbf{r}$ simultaneously, i.e., the elements 
$g$ and $g'$ must be the same for all $\mathbf{r}$.
\item For a given irrep $\mathbf{r}$, the consistency condition defines $X_\mathbf{r}$ 
up to an overall phase and a $G_f$ transformation.
\item It follows from eq.~\eqref{eq:consistency} that the elements $g$ and $g'$ must be of the same order.
\item It is sufficient to impose eq.~\eqref{eq:consistency} on the generators of a discrete group $G_f$.
\item The chain $CP \to g \to CP^{-1}$
maps the group element $g$ onto $g'$ and
preserves the flavour symmetry group structure. Therefore,  
it realises a homomorphism $v(g) = g'$ of $G_f$.
Assuming the presence of faithful representations $\mathbf{r}$,
i.e., those for which $\rho_\mathbf{r}$ 
maps each element of $G_f$ to a distinct matrix,
eq.~\eqref{eq:consistency} 
defines a unique mapping of $G_f$ to itself. 
In this case, $v(g)$ is an automorphism of $G_f$.
\item The automorphism $v(g) = g'$ must be class-inverting 
with respect to $G_f$,
i.e.~$g'$ and $g^{-1}$ belong to the same conjugacy 
class~\cite{Chen:2014tpa}. It is furthermore an outer automorphism,
meaning no $h \in G_f$ exists such that $g' = h^{-1}gh$.
\end{itemize}
%%%%%%%
%

 It has been shown in Ref.~\cite{Feruglio:2012cw} that 
under the assumption of $X_\mathbf{r}$ being a symmetric matrix,%
\footnote{In this case, the CP transformation applied twice to a 
chiral superfield gives the field itself 
(see eq.~\eqref{eq:GCPtransformation}).
We note also that the CP transformation applied twice to a fermion field gives the field multiplied by a minus sign, which is consistent with the superfield transformation (see Appendix~\ref{app:SUSYCP}).
} 
the full symmetry group is isomorphic to a semi-direct product 
$G_f \rtimes H_{CP}$, where $H_{CP} \simeq \mathbb{Z}_2^{CP}$
is the group generated by the gCP transformation.

Finally, we would like to note that for $G_f = S_3$, $A_4$, $S_4$ and $A_5$
and in the bases for their representation matrices summarised
in Appendix~\ref{app:sym_basis}, 
the gCP transformation $X_\mathbf{r} = \mathbbm{1}_\mathbf{r}$ 
up to inner automorphisms, i.e, $X_\mathbf{r} = \rho_\mathbf{r}(g),~g \in G_f$, 
as shown in Refs.~\cite{Ding:2013bpa}, \cite{Holthausen:2012dk}
and \cite{Li:2015jxa}.

%=======================
%
\subsection{Modular Symmetry and Modular-Invariant Theories}
\label{sec:modular_sym}
%
%=======================
%
In this subsection, we briefly summarise the modular invariance approach 
to flavour~\cite{Feruglio:2017spp}.
An element $\gamma$ of the modular group $\overline\Gamma$
acts on a complex variable $\tau$ belonging to the upper-half 
complex plane as follows:
%%%%%%%
\be
\g\t = \frac{a\t + b}{c\t + d}\,,
\quad
\text{where} 
\quad
a,b,c,d \in \mathbb{Z}
\quad
\text{and}
\quad
ad - bc = 1\,,~~{\rm Im}\tau > 0\,.
\label{eq:linfractransform}
\ee
%%%%%%%
%
 The modular group $\overline\Gamma$ is isomorphic to 
the projective special linear group 
$PSL(2,\mathbb{Z}) = SL(2,\mathbb{Z})/\mathbb{Z}_2$,
where  $SL(2,\mathbb{Z})$ is the special linear group 
of integer $2\times 2$ matrices with unit determinant 
and $\mathbb{Z}_2 = \{I,-I\}$ is its centre, $I$ being the identity element.
The group  $\overline\Gamma$
can be presented in terms of two generators $S$ and $T$ satisfying
%%%%%%%
%
\be
S^2 = \left(ST\right)^3 = I\,.
\ee
%%%%%%%
The generators admit the following matrix representation:
%%%%%%%
\be
S = \begin{pmatrix}
0 & -1 \\
1 & 0
\end{pmatrix}\,,
\qquad
T = \begin{pmatrix} 
1 & 1 \\
0 & 1
\end{pmatrix}\,.
\ee
%%%%%%%
%
The action of $S$ and $T$ on $\tau$ amounts to inversion 
with a change of sign and translation, respectively:
%%%%%%%
\be
\t \xrightarrow{S} -\frac{1}{\t}\,, 
\qquad
\t \xrightarrow{T} \t + 1\,.
\ee
%%%%%%%
%

Let us consider the infinite normal subgroups 
$\G(N)$, $N = 1,2,3,\dots$, of $SL(2,\mathbb{Z})$
(called also the principal congruence subgroups):
%%%%%
\be
\G(N) = \left\{
\begin{pmatrix}
a & b \\
c & d
\end{pmatrix} 
\in SL(2,\mathbb{Z})\,, 
\quad
\begin{pmatrix}
a & b \\
c & d
\end{pmatrix} =
\begin{pmatrix}
1 & 0 \\
0 & 1
\end{pmatrix} 
~~(\text{mod } N)
\right\}.
\ee
%%%%%
%
For $N=1$ and $2$, one defines the groups
$\overline{\G}(N) \equiv \G(N)/\{I,-I\}$ 
(note that $\overline{\G}(1) \equiv \overline{\G}$), 
while for $N > 2$, $\overline{\G}(N) \equiv \G(N)$. 
The quotient groups $\G_N \equiv \overline{\G}/\overline{\G}(N)$
turn out to be finite. They are referred to as finite modular groups. 
Remarkably, for $N \leq 5$, these groups are isomorphic to 
permutation groups:
$\G_2 \simeq S_3$, $\G_3 \simeq A_4$, $\G_4 \simeq S_4$ and $\G_5 \simeq A_5$.
Their group theory is summarised in Appendix~\ref{app:group}.
We recall here only that the group $\G_N$ 
is presented by two generators $S$ and $T$ satisfying:
%%%%%%%%%%%%%
\be 
 S^2 = (ST)^3 =  T^N = I\,. 
\label{eq:GammaNgener}
\ee
%%%%%%%%%%%%%%%%%%%%%%%%
We will work in the basis in which the generators $S$ and $T$ of these groups are 
represented by symmetric matrices,
%%%%%%%%%%%%%%%%%%%%%%%%%%%%%
\begin{equation}
  \rho_\mathbf{r}(S) = \rho^T_\mathbf{r}(S)\,, \quad
  \rho_\mathbf{r}(T) = \rho^T_\mathbf{r}(T)\,,
\label{eq:sym_generators}
\end{equation}
%%%%%%%%%%%%%%%%%%%%%%%%%%%%%%%
%
for all irreducible representations $\mathbf{r}$.
The convenience of this choice will become clear later on. 
For the groups $\Gamma_N$ with $N \leq 5$, the working bases 
are provided in Appendix~\ref{app:sym_basis}.

 The key elements of the considered framework 
 are modular forms $f(\t)$ of weight $k$ and level $N$. 
These are holomorphic functions, which transform under 
$\overline{\G}(N)$ as follows:
%%%%%
\be
f\left(\g\t\right) = \left(c\t + d\right)^k f(\t)\,, 
\quad 
\g \in \overline{\G}(N)\,,
\ee
%%%%%
%
where the weight $k$ is an even and non-negative number, 
and the level $N$ is a natural number.
For certain $k$ and $N$, the modular forms
span a linear space of finite dimension.
One can find a basis in this space such that 
a multiplet of modular forms $F(\t) \equiv (f_1(\t), f_2(\t), \dots)^T$ 
transforms according to a unitary representation $\mathbf{r}$ of $\G_N$:
%%%%%
\be
F\left(\g\t\right) = \left(c\t + d\right)^k \rho_\mathbf{r}(\g) F(\t)\,, 
\quad 
\g \in \overline{\G}\,.
\label{eq:vvmodforms}
\ee
%%%%%
%
In Appendix~\ref{app:lw_multiplets}, we provide the multiplets 
of modular forms of 
lowest non-trivial weight $k=2$ at levels $N = 2,3,4$ and $5$, 
i.e., for $S_3$, $A_4$, $S_4$ and $A_5$.
Multiplets of higher weight modular forms can be constructed 
from tensor products of the lowest weight multiplets. 
For $N=4$ (i.e., $S_4$), we present in Appendix~\ref{app:higher_weights} 
modular multiplets of weight $k \leq 10$ derived in the symmetric basis 
for the $S_4$ generators (see Appendix~\ref{app:sym_basis}). 
For $N=3$ and $N=5$ (i.e., $A_4$ and $A_5$), modular multiplets of weight 
up to $6$ and $10$, computed in the bases employed by us,
can be found in \cite{Feruglio:2017spp} and \cite{Novichkov:2018nkm}, 
respectively.

 In the case of $\mathcal{N} = 1$ rigid SUSY, 
the matter action $\mathcal{S}$ reads
%%%%%
\be
\mathcal{S} = \int \di^4x\, \di^2\th\, \di^2\overline{\th}~ 
K(\t, \overline{\t}, \psi, \overline{\psi}) + 
\int \di^4x\, \di^2\th~W(\t, \psi) + 
\int \di^4x\, \di^2\overline{\th}~\overline{W}(\overline{\t}, \overline{\psi})\,,
\label{eq:SUSYaction}
\ee
%%%%%
%
where $K$ is the Kähler potential, $W$ is the superpotential, 
$\psi$ denotes a set of chiral supermultiplets $\psi_i$, 
and $\tau$ is the modulus chiral superfield, whose lowest 
component is the complex scalar field acquiring a VEV.%
\footnote{We will use the same notation $\tau$ 
for the lowest complex scalar component of the modulus superfield 
and will call this component also ``modulus'' since 
in what follows we will be principally concerned 
with this scalar field.}
$\th$ and $\overline{\th}$ are Graßmann variables.
The modulus $\t$ and supermultiplets $\psi_i$  
transform under the action of the modular group 
in a certain way~\cite{Ferrara:1989bc,Ferrara:1989qb}.
Assuming, in addition, that the supermultiplets $\psi_i = \psi_i(x)$ transform 
in a certain irreducible representation $\mathbf{r}_i$ of $\G_N$, 
the transformations read:
%%%%%
\be
\g = \begin{pmatrix}
 a & b  \\ c & d
\end{pmatrix}
\in \overline\G :\quad
\begin{cases}
\t \rightarrow \dfrac{a\t + b}{c\t + d}\,, \\[4mm]
\psi_i \rightarrow \left(c\t + d\right)^{-k_i} \rho_{\mathbf{r}_i}(\g)\, \psi_i\,.
\end{cases}
\label{eq:modtransforms}
\ee
%%%%%
%
It is worth noting that $\psi_i$ is not a multiplet of modular forms, 
and hence, the weight $(- k_i)$ can be 
positive or negative, and can be even or odd.
Invariance of the matter action under these transformations implies
%%%%%
\be
\begin{cases}
W(\t, \psi) \rightarrow W(\t,\psi)\,, \\[4mm]
K(\t, \overline{\t}, \psi, \overline{\psi}) \rightarrow K(\t, \overline{\t}, \psi, \overline{\psi}) 
+ f_K(\t,\psi) + \overline{f_K}(\overline{\t},\overline{\psi})\,,
\end{cases}
\ee
%%%%%
%
where the second line represents a Kähler transformation.\\

An example of the Kähler potential, which we will use
in what follows, reads:
%%%%%%%%%%%%%%%%%%%%%%%%%%%
\be
K(\t, \overline{\t}, \psi, \overline{\psi})
= - \Lambda_0^2 \log(-i\tau + i \overline{\tau})
  + \sum_i \frac{|\psi_i|^2}{(-i\tau + i \overline{\tau})^{k_i}} \,, 
\label{eq:Kahler}
\ee
%%%%%%%%%%%%%%%%%%%%%%%%%%%%%%
%
with $\Lambda_0$ having mass dimension one.
The superpotential can be expanded in powers of $\psi_i$ as follows:
%%%%%
\be
W(\t, \psi) = \sum_{n} \sum_{\{i_1,\dots,i_n\}}
\sum_s \,g_{i_1\,\dots\,i_n,s} 
 \left(Y_{i_1\,\dots\,i_n, s}(\t)\, \psi_{i_1}\dots\psi_{i_n}\right)_{\1,s}\,,
\label{eq:superpotentialGen}
\ee
%%%%%
%
where $\1$ stands for an invariant singlet of $\G_N$.
For each set of $n$ fields $\{\psi_{i_1},\dots,\psi_{i_n}\}$,
the index $s$ labels the independent singlets.
Each of these is accompanied by a coupling 
constant $g_{i_1\,\dots\,i_n,s}$ and is obtained using 
a modular multiplet $Y_{i_1\,\dots\,i_n, s}$ of the requisite weight.
Indeed, to ensure invariance of $W$ under the transformations 
in eq.~\eqref{eq:modtransforms}, 
the set $Y_{i_1\,\dots\,i_n,s}(\t)$ of functions must transform 
in the following way (we omit indices for brevity):
%%%%%
\be
Y(\t) \,\xrightarrow{\g}\, (c\t + d)^{k_{Y}} \rho_{\mathbf{r}_Y}(\g)\, Y(\t)\,,
\label{eq:multiplet_trans}
\ee
%%%%%
%
where $\mathbf{r}_{Y}$ is a representation of 
$\G_N$, and $k_{Y}$ and $\mathbf{r}_{Y}$ 
are such that
%%%%%
\begin{align}
&k_{Y} = k_{i_1} + \dots + k_{i_n}\,, \\[2mm]
&\mathbf{r}_{Y} \ot \mathbf{r}_{i_1} \ot \dots \ot \mathbf{r}_{i_n} \supset \1\,.
\end{align}
%%%%%
%
Thus, $Y_{i_1\,\dots\,i_n,s}(\t)$ represents a multiplet 
of weight $k_{Y}$ and level $N$ modular forms 
transforming in the representation $\mathbf{r}_{Y}$ of $\G_N$
(cf.~eq.~\eqref{eq:vvmodforms}).

%%%%%%%%%%%%%%%%%%%%%%%
%
\section{gCP Transformations Consistent with Modular Symmetry}
\label{sec:modularGCP}
%
%%%%%%%%%%%%%%%%%%%%%%%
%
As we saw in subsection~\ref{sec:gcp_general}, CP transformations
can in general be combined with flavour symmetries in a non-trivial way.
In the set-up of subsection~\ref{sec:modular_sym}, the role of flavour symmetry 
is played by modular symmetry.
In this section, we derive the most general form of a CP transformation 
consistent with modular symmetry.
Unlike the case of discrete flavour symmetries, field transformation 
properties under CP are restricted to a unique possibility,
given the transformation of the modulus (see subsection~\ref{sec:tauCP}) 
and eq.~\eqref{eq:GCPtransformation}.
The derivation we are going to present is
agnostic to the UV completion of the theory and, 
in particular, the origin of modular symmetry.

%=======================
\subsection{CP Transformation of the Modulus \texorpdfstring{$\tau$}{τ}}
\label{sec:tauCP}
%=======================
%
Let us first apply the consistency condition chain%
\footnote{
It may be possible to generalise the CP transformation
such that it can be combined not only with modular
but also with other internal symmetries of the theory.
We are not going to consider this case here.
}
%%%%%%%%%%%%%%%%%%%%%
\be 
CP \,\to\, \gamma\, \to CP^{-1} \,=\, \gamma' \in \overline\Gamma
\label{eq:cons_cond}
\ee
%%%%%%%%%%%%%%%
%
to an arbitrary chiral superfield $\psi(x)$ 
assigned to an irreducible unitary representation $\mathbf{r}$ 
of $\Gamma_N$, which transforms as
$\psi(x) \to X_{\mathbf{r}}\, \overline\psi(x_P)$ under CP:
%%%%%%%%%%%%%%%%%%%%%%%%%%%
\begin{equation}
\begin{aligned}
 \psi(x) &\xrightarrow{CP}
 X_\mathbf{r} \,\overline\psi(x_P) \xrightarrow{\gamma}
 (c\tau^{*} + d)^{-k} \, X_\mathbf{r} \,\rho_\mathbf{r}^{*}(\gamma) \, \overline\psi(x_P) \\
 &\xrightarrow{CP^{-1}}
 (c\tau^{*}_{CP^{-1}} + d)^{-k} \, X_\mathbf{r} \,\rho_\mathbf{r}^{*}(\gamma) X_\mathbf{r}^{-1} \, \psi(x)\,,
\end{aligned}
\end{equation}
%%%%%%%%%%%%%%%%%%%%%%%%%%%%%%
%
where $\tau_{CP^{-1}}$ is the result of applying $CP^{-1}$ to the modulus $\tau$.
The resulting transformation should be equivalent to a modular 
transformation $\gamma'$ which depends on $\gamma$ and maps $\psi(x)$ 
to $(c'\tau + d')^{-k} \rho_\mathbf{r}(\gamma') \psi(x)$. 
Taking this into account, we get
%%%%%%%%%%%%%%%%%%%%%%%%%%%%%%%%%%%%%%%%%
\begin{equation}
  X_\mathbf{r} \,\rho_\mathbf{r}^{*}(\gamma) X_\mathbf{r}^{-1} = \left( \frac{c'\tau + d'}{c\tau^{*}_{CP^{-1}} + d} \right)^{-k} \rho_\mathbf{r}(\gamma') \, .
  \label{eq:Xcond_prelim}
\end{equation}
%%%%%%%%%%%%%%%%%%%%%%%%%%%%%
%
Since the matrices $X_\mathbf{r}$,
$\rho_\mathbf{r}(\gamma)$ and $\rho_\mathbf{r}(\gamma')$ are
independent of $\tau$, the overall coefficient on the right-hand
side has to be a constant:%
\footnote{
  Strictly speaking, this is only true for non-zero weights $k$.
  We assume that at least one superfield with non-zero modular weight exists 
in the theory, because otherwise the modulus has no effect on the 
superfield transformations, and the modular symmetry approach reduces 
to the well-known discrete symmetry approach to flavour.
}
%%%%%%%%%%%%%%%%%%%%%%%%%%%%%%%%%%%
\begin{equation}
  \frac{c'\tau + d'}{c\tau^{*}_{CP^{-1}} + d} = \frac{1}{\lambda^{*}} \, ,
  \label{eq:coef_lambda}
\end{equation}
%%%%%%%%%%%%%%%%%%%%%%%%%%%%%%
%
where $\lambda \in \mathbb{C}$, and $|\lambda| = 1$ due to unitarity of 
 $\rho_\mathbf{r}(\gamma)$ and $\rho_\mathbf{r}(\gamma')$.
The values of $\lambda$, $c'$ and $d'$ depend on $\gamma$.

Taking $\gamma = S$, so that $c = 1$, $d = 0$, and denoting
$c'(S) = C$, $d'(S) = D$
while keeping henceforth the notation $\lambda(S) = \lambda$, 
we find $\tau = (\lambda\tau^{*}_{CP^{-1}} - D)/C$, and 
consequently,
%%%%%%%%%%%%%%%%%%%%%%%%%%%%%%%%
\begin{equation}
  \tau \,\xrightarrow{CP^{-1}}\, \tau_{CP^{-1}} =  
\lambda \left( C\tau^{*} + D \right), \quad
  \tau \,\xrightarrow{CP}\,  \tau_{CP} =   
\frac{1}{C} \left( \lambda \tau^{*} - D \right) \, .
  \label{eq:taucp_prelim}
\end{equation}
%%%%%%%%%%%%%%%%%%%%%%%%%%%%%%%
%
Let us now act with the chain $CP \to T \to CP^{-1}$ on the modulus 
$\tau$ itself:
%%%%%%%%%%%%%%%%%%%%%%%%%%%%%%
\begin{equation}
  \tau\, \xrightarrow{CP}\, \frac{1}{C} \left( \lambda \tau^{*} - D \right)
  \,\xrightarrow{T}\, \frac{1}{C} \left( \lambda \left( \tau^{*} + 1 \right) - D \right)
  \,\xrightarrow{CP^{-1}}\, \tau + \frac{\lambda}{C} \, .
\end{equation}
%%%%%%%%%%%%%%%%%%%%%%
%
The resulting transformation has to be a modular transformation, 
therefore $\lambda / C \in \mathbb{Z}$.
Since $|\lambda| = 1$, we immediately find $|C| = 1$, $\lambda = \pm 1$.
After choosing the sign of $C$ as $C = \mp 1$ so that $\im \tau_{CP} > 0$, 
the CP transformation rule~\eqref{eq:taucp_prelim} simplifies to
%%%%%%%%%%%%%%%%%%%%%%%%%%%%%%%
\begin{equation}
  \tau \,\xrightarrow{CP}\, n - \tau^{*}\,,
  \label{eq:taucp_shift}
\end{equation}
%%%%%%%%%%%%%%%%%%%%%%%
%
with $n \in \mathbb{Z}$.
One can easily check that the chain 
$CP \to S \to CP^{-1} = \gamma'(S)$ 
(applied to the modulus $\tau$ itself) imposes no further 
restrictions on the form of $\tau_{CP}$.
Since $S$ and $T$ generate the entire modular group, we conclude 
that eq.~\eqref{eq:taucp_shift} is the most general CP transformation 
of the modulus $\tau$ compatible with the modular symmetry.

It is always possible to redefine the CP transformation 
in such a way that $n = 0$.
Consider the composition 
$CP' \equiv T^{-n} \circ CP$ so that $\tau \xrightarrow{CP'} - \tau^{*}$.
It is worth noting that this redefinition represents 
an inner automorphism which does not spoil 
the form of gCP transformation in eq.~\eqref{eq:GCPtransformation}.
Indeed, the chiral superfields transform under $CP'$ as
%%%%%%%%%%%%%%%%%%%%%%%%%%%%%%%
\begin{equation}
  \psi \,\xrightarrow{CP'}\, \rho^{-n}_\mathbf{r}(T)\, X_\mathbf{r}\, \overline{\psi} \,.
\end{equation}
%%%%%%%%%%%%%%%%%%%%%%%%%%%%%%%%%
%
Thus, $CP'$ has the same properties as the original CP transformation
up to a redefinition of $X_\mathbf{r}$.
Therefore, from now on we will assume without loss of generality 
that the modulus $\tau$ transforms under CP as%
\footnote{
The CP transformation of the modulus derived by us
from the requirement of consistency between
modular and CP symmetries has appeared
in the context of string-inspired models
(see, e.g., Refs.~\cite{Acharya:1995ag, Dent:2001cc, Giedt:2002ns, Baur:2019kwi}).
}
%%%%%%%%%%%%%%%%%%%%%%%%%%%%%%%%%%%%
\begin{equation}
  \label{eq:taucp}
  \tau \,\xrightarrow{CP}\, - \tau^{*} \, .
\end{equation}
%%%%%%%%%%%%%%%%%%%%%%%%%%%%%%%%%%%%
% 
It obviously follows from the preceding equation that 
$\tau$ does not change under the action of $CP^2$: 
%%%%%%%%%%%%%%%%%%%%%%%%%%%%%%%%%%%%
\begin{equation}
  \label{eq:taucp2}
  \tau \,\xrightarrow{CP^2}\, \tau \,.
\end{equation}
%%%%%%%%%%%%%%%%%%%%%%%%%%%%%%%%%%%%
%
Thus, in what concerns the action on the modulus $\tau$ 
we have: $CP^2 = I$, $CP^{-1} = CP$.

%=======================
\subsection{Extended Modular Group}
\label{sec:ext_mod_group}
%=======================
%
Having derived the explicit form of the CP transformation for 
the modulus $\tau$, we are now in a position to find the action of CP 
on the modular group $\overline{\Gamma}$ as an outer automorphism 
$u(\gamma)$.
For any modular transformation $\gamma \in \overline{\Gamma}$ we have
%%%%%%%%%%%%%%%%%%%%%%%%%%
\begin{equation}
  \tau \,\xrightarrow{CP}\, -\tau^{*}
  \,\xrightarrow{\gamma}\, - \frac{a\tau^{*} + b}{c\tau^{*} + d}
  \,\xrightarrow{CP^{-1}}\, \frac{a\tau - b}{-c\tau + d} \,.
\end{equation}
%%%%%%%%%%%%%%%%%%%%%%%%%%%%%
%
This implies that the sought-after automorphism is
%%%%%%%%%%%%%%%%%%%%%%%%%%%%%%%
\begin{equation}
  \label{eq:conj_aut}
  \gamma =
  \begin{pmatrix}
    a & b \\
    c & d
  \end{pmatrix}
  \,\to\,\,
  u(\gamma) \,\equiv\, CP\, \gamma\, CP^{-1} \,=\, 
  \begin{pmatrix}
    a & -b \\
    -c & d
  \end{pmatrix} \,.
\end{equation}
%%%%%%%%%%%%%%%%%%%%%%%%%%%%%%%%%%%
%
In particular, one has 
$(CP) \, S \, (CP)^{-1} = S$, $(CP)\, T\, (CP)^{-1} = T^{-1}$, or simply
$u(S) = S$ and $u(T) = T^{-1}$.
It is straightforward to check that the mapping~\eqref{eq:conj_aut} 
is indeed an outer automorphism of $\overline{\Gamma}$.%
\footnote{One can explicitly check that 
i) $u(\gamma_1)u(\gamma_2) = u(\gamma_1\gamma_2)$, 
meaning $u$ is an automorphism, and that 
ii) there is no group element $\widetilde\gamma \in \overline\Gamma$  
such that $u(\gamma) = \widetilde\gamma^{-1} \gamma\, \widetilde\gamma$, 
meaning that $u$ is an outer automorphism.}
Notice further that if $\gamma \in \overline\Gamma(N)$, then also $u(\gamma) \in \overline\Gamma(N)$.

By adding the CP transformation~\eqref{eq:taucp} as a new generator 
to the modular group, one obtains the so-called {\it extended modular group}:
%%%%%%%%%%%%%%%%%%%%%%%%%%%
\begin{equation}
  \overline{\Gamma}^{*}
  = \left\langle \tau \xrightarrow{T} \tau + 1, \, \tau \xrightarrow{S} -1/\tau, \, \tau \xrightarrow{CP} -\tau^{*} \right\rangle
\end{equation}
%%%%%%%%%%%%%%%%%%%%%%%%%%%%%%%%%%%%%
%
(see, e.g., \cite{Kulkarni:1991}), which has a structure of 
a semi-direct product 
$\overline{\Gamma}^{*} \simeq \overline{\Gamma} \rtimes \mathbb{Z}_2^{CP}$, 
with $\mathbb{Z}_2^{CP} = \{I, \tau \to -\tau^{*}\}$.
The group $\overline{\Gamma}^{*}$ is isomorphic to the group 
$PGL(2,\mathbb{Z})$ of integral 
$2 \times 2$ matrices with determinant $\pm 1$, 
the matrices $M$ and $-M$ being identified.
The CP transformation is then represented by the matrix
%%%%%%%%%%%%%%%%%%%%%%%%%%%%%%%%
\begin{equation}
  CP =  \begin{pmatrix} 1 & 0 \\ 0 & -1 \end{pmatrix}\,, 
\quad CP\,\gamma\, CP^{-1} = u(\gamma)\,.
\label{eq:CPmatrix}
\end{equation}
%%%%%%%%%%%%%%%%%%
%
The action of $\overline{\Gamma}^{*}$ on the complex 
upper-half plane is defined as
%%%%%%%%%%%%%%%%%%%%%%%%%%%%%%
\begin{equation}
  \begin{pmatrix} a & b \\ c & d \end{pmatrix} 
  \in \overline\G^*:\quad
  \begin{cases}
    \tau \to \cfrac{a\tau + b}{c\tau + d} &\quad \text{if} \quad ad - bc = 1\,, \\
    \tau \to \cfrac{a\tau^{*} + b}{c\tau^{*} + d} 
&\quad \text{if} \quad ad - bc = -1\,.
  \end{cases}
\end{equation}
%%%%%%%%%%%%%%%%%%%%%%%%%%%%%%%%%%

%=======================
\subsection{CP Transformation of Chiral Superfields}
\label{sec:CPchsf}
%=======================

A chiral superfield $\psi(x)$
transforms according to eq.~\eqref{eq:GCPtransformation} under CP.
The consistency condition chain~\eqref{eq:cons_cond} 
applied to $\psi$ constrains the form of its CP transformation
matrix $X_\mathbf{r}$ 
as in eq.~\eqref{eq:Xcond_prelim}, with the overall coefficient
on the right-hand side
being constant, as discussed earlier, see eq.~\eqref{eq:coef_lambda}.
Since $\lambda = \pm 1$, the coefficient on the right-hand side
of eq.~\eqref{eq:Xcond_prelim} is $(\pm 1)^k$.
This sign is actually determined by the signs of matrices 
in the outer automorphism~\eqref{eq:conj_aut}, which 
are unphysical in a modular symmetric theory.
By choosing%
\footnote{
This choice is possible since
in the presence of fields with an odd modular weight
the theory automatically respects a sign-flip $\mathbb{Z}_2$ symmetry, 
which acts only non-trivially on these fields.
This implies that the discussed sign is unphysical.}
these signs in such a way that $c' = -c$, $d' = d$, 
in accordance with eq.~\eqref{eq:conj_aut},
we obtain a trivial coefficient $+1$,
and the constraint on $X_\mathbf{r}$ reduces to
%%%%%%%%%%%%%%%%%%%%%%%
\begin{equation}
  \label{eq:Xcond}
  X_\mathbf{r}\, \rho_\mathbf{r}^{*}(\gamma) X_\mathbf{r}^{-1} = \rho_\mathbf{r}(\gamma') \,.
\end{equation}
%%%%%%%%%%%%%%%%%%%%%%%%%%%%%%%%%
%
The constraint we get coincides with the corresponding
constraint in the case of non-Abelian discrete flavour symmetries, 
eq.~\eqref{eq:consistency}.
However, unlike in the usual discrete flavour symmetry approach, 
modular symmetry restricts the form of the automorphism 
$\gamma \to \gamma' = u(\gamma)$ to the unique possibility 
given in eq.~\eqref{eq:conj_aut},
which acts on the generators as $S \to u(S) = S$ and $T \to u(T) = T^{-1}$.
Therefore, for each irreducible representation
$\mathbf{r}$, $X_\mathbf{r}$ in eq.~\eqref{eq:Xcond} is fixed up to
an overall phase by Schur's lemma.

For the working bases discussed in subsection~\ref{sec:modular_sym}
and given in Appendix~\ref{app:sym_basis}, 
one has $X_\mathbf{r} =  \mathbbm{1}_\mathbf{r}$, i.e.,
the gCP transformation has the canonical form.
The key feature of the aforementioned bases which allows for this
simplification is that the group generators $S$ and $T$ are
represented by symmetric matrices.
Indeed, if eq.~\eqref{eq:sym_generators} holds, one has:
%%%%%%%%%%%%%%%%%%%%%%%%%%%%%%%%%%%%
\begin{equation}
  \rho^{*}_\mathbf{r}(S) = \rho^{\dagger}_\mathbf{r}(S) = 
\rho_\mathbf{r}(S^{-1}) = \rho_\mathbf{r}(S)\,, \quad
  \rho^{*}_\mathbf{r}(T) = \rho^{\dagger}_\mathbf{r}(T) 
= \rho_\mathbf{r}(T^{-1})\,,
\end{equation}
%%%%%%%%%%%%%%%%%%%%%%%%%%%%
%
so that $X_\mathbf{r} =  \mathbbm{1}_\mathbf{r}$ solves the consistency 
condition~\eqref{eq:Xcond}.

%=======================
\subsection{CP Transformation of Modular Form Multiplets}
 \label{sec:multCP}
%=======================

Since modular multiplets $Y(\tau)$ transform under the modular group 
in essentially the same way as chiral superfields,
it is natural to expect that the above discussion holds for modular 
multiplets as well.
In particular, they should transform under CP as $Y \to X_\mathbf{r}\, Y^{*}$.
Still, it is instructive to derive their transformation rule explicitly.

Under a modular transformation, $Y(\tau)$ transforms as in 
eq.~\eqref{eq:multiplet_trans},
while under the action of CP one has $Y(\tau) \to Y(-\tau^*)$.
It can be shown (see Appendix D of~\cite{Novichkov:2018ovf}) that
the complex-conjugated CP-transformed multiplets $Y^*(-\tau^*)$
transform almost like the original multiplets $Y(\tau)$ under a modular transformation, namely:
%%%%%%%%%%%%%%%%%%%%%%%%%%%%%%%%%%%%%%%%
\begin{equation}
  Y^*(-\tau^*) \, \xrightarrow{\gamma}\, Y^*\left(-(\gamma\tau)^*\right) \,=\, 
Y^*\left(u(\gamma)(-\tau^*)\right)  \,=\,
(c\tau + d)^{k}\, \rho^*_\mathbf{r}\left(u(\gamma)\right)\, Y^*(-\tau^*)\,,
\end{equation}
%%%%%%%%%%%%%%%%%%%%%%%%%%%
%
for a multiplet $Y(\tau)$ of weight $k$ transforming in the 
irreducible representation $\mathbf{r}$ of $\Gamma_N$.

Using the consistency condition in eq.~\eqref{eq:Xcond}, one then sees that
it is the object $X_{\mathbf{r}}^T\, Y^*(-\tau^*)$ which transforms like $Y(\tau)$ under a modular transformation,
i.e.:
\begin{equation}
  X_{\mathbf{r}}^T \,Y^*(-\tau^*) \, \xrightarrow{\gamma}\, (c\tau + d)^{k}\, \rho_\mathbf{r}(\gamma) \, \left[X_{\mathbf{r}}^T\, Y^*(-\tau^*)\right]\,.
\end{equation}

If there exists a unique modular form multiplet at a certain level $N$, weight $k$ and representation $\mathbf{r}$, then proportionality follows:
%%%%%%%%%%%%%%%%%%%%%%%%%%
\begin{equation}
  \label{eq:prop}
  Y(\tau) \,=\, z\,X_{\mathbf{r}}^T \,Y^*(-\tau^*) \,,
\end{equation}
%%%%%%%%%%%%%%%%%%%%%%
%
with $z \in \mathbb{C}$.
This is indeed the case for $2 \leq N \leq 5$ and lowest weight $k = 2$.
Since $Y\left(-(-\tau^*)^*\right) = Y(\tau)$, it follows that $X_\mathbf{r} \, X_\mathbf{r}^* = |z|^2  \mathbbm{1}_\mathbf{r}$,
implying i) that $z= e^{i\phi}$ is a phase which can be absorbed in the normalisation of $Y(\tau)$, and ii) that
$X_\mathbf{r}$ must be symmetric in this case, $X_\mathbf{r}\, X_\mathbf{r}^* =  \mathbbm{1}_\mathbf{r} \Rightarrow X_\mathbf{r} = X_\mathbf{r}^T$,
independently of the basis.
One can then write
%%%%%%%%%%%%%%%%%%%%%%%%%%%%%%%%%
\begin{equation}
    Y(\tau) \,\xrightarrow{CP}\, Y(-\tau^*) = X_\mathbf{r} \,Y^*(\tau)
    \label{eq:ycp}
\end{equation}
%%%%%%%%%%%%%%%%%%%%%%%%%
%
for these multiplets, as anticipated.

As we have seen in subsection~\ref{sec:CPchsf}, 
in a basis in which the generators $S$ and $T$ of $\Gamma_N$ are represented by symmetric matrices,
one has $X_\mathbf{r} = \mathbbm{1}_\mathbf{r}$. From eq.~\eqref{eq:prop} it follows that $Y(-\tau^{*}) = e^{i\phi}\, Y^*(\tau)$,
the phase $\phi$ being removable, as commented above.
At the $q$-expansion level this means that, in such a basis, all the expansion coefficients are real up to a common complex phase.
This is indeed the case for the lowest-weight modular form multiplets of $\Gamma_N$ with $N\leq 5$,
as can be explicitly verified from the $q$-expansions 
collected in Appendix~\ref{app:qexpansions}.
This is further the case for the higher-weight modular 
multiplets of these groups in such a basis,
given the reality of Clebsch-Gordan coefficients, 
summarised in Appendix~\ref{app:CGCs}.

%%%%%%%%%%%%%%%%%%%%%%%
\section{CP-Invariant Theories with Modular Symmetry}
\label{sec:CP_inv}
%%%%%%%%%%%%%%%%%%%%%%%

%%%%%%%%%%%%%%%%%%%%%%%
\subsection{Implications of CP Invariance for the Couplings}
%%%%%%%%%%%%%%%%%%%%%%%

We have found so far that a CP transformation consistent with modular symmetry acts on fields and modular form multiplets in the following way:
%%%%%%%%%%%%%%%%%%%%%%%%%
\begin{equation}
  \tau \,\xrightarrow{CP}\, -\tau^*, \quad
  \psi(x) \,\xrightarrow{CP}\, X_{\mathbf{r}} \,\overline\psi(x_P), \quad
  Y(\tau) \,\xrightarrow{CP}\, Y(-\tau^*) = X_{\mathbf{r}}\, Y^{*}(\tau) \, .
  \label{eq:GCPall}
\end{equation}
%%%%%%%%%%%%%%%%%%%%%
%
A SUSY modular-invariant theory is thus CP-conserving if the transformation~\eqref{eq:GCPall} leaves the matter action $\mathcal{S}$ given by eq.~\eqref{eq:SUSYaction} unchanged.
In particular, the superpotential $W$ has to transform into its Hermitian conjugate, while the Kähler potential $K$ is allowed to change by a Kähler transformation.

The Kähler potential of eq.~\eqref{eq:Kahler} is clearly invariant under the CP transformation~\eqref{eq:GCPall}, since it depends on $|\psi|^2$ and $\im \tau$, both of which remain unchanged (up to a change $x\to x_P$ 
which does not affect $\mathcal{S}$). 
On the other hand, the superpotential can be written as a sum of independent terms of the form
%%%%%%%%%%%%%%%%%%%%%%%
\begin{equation}
 W \supset
 \sum_s g_s \left( Y_s(\tau) \psi_1 \ldots \psi_n \right)_{\mathbf{1},s} \, ,
  \label{eq:Wterm}
\end{equation}
%%%%%%%%%%%%%%%%%%%%%
%
where $Y_s(\tau)$ are modular multiplets of a certain weight and irreducible representation,
and $g_s$ are complex coupling constants, see eq.~\eqref{eq:superpotentialGen}.
Such terms transform non-trivially under CP, which leads to a certain constraint on the couplings $g_s$.

This can be easily checked for a symmetric basis, as in this basis $X_{\mathbf{r}} = \mathbbm{1}_\mathbf{r}$
for any representation $\mathbf{r}$, so that one has (assuming proper normalisation of the modular multiplets $Y_s(\tau)$)
%%%%%%%%%%%%%%%%%%%%%%%%%%%
\begin{equation}
\begin{aligned}
  g_s \left( Y_s(\tau) \psi_1 \ldots \psi_n \right)_{\mathbf{1},s} 
  &\xrightarrow{CP}
  g_s \left( Y_s(-\tau^{*}) \overline\psi_1 \ldots \overline\psi_n \right)_{\mathbf{1},s} \\
  &\quad= g_s \left( Y_s^{*}(\tau) \overline\psi_1 \ldots \overline\psi_n \right)_{\mathbf{1},s} =
  g_s \,\overline{\left( Y_s(\tau) \psi_1 \ldots \psi_n \right)_{\mathbf{1},s}} \, ,\label{eq:CPWterm}
\end{aligned}
\end{equation}
%%%%%%%%%%%%%%%%%%%%%%%
%
where in the last equality we have used the reality of the 
Clebsch-Gordan coefficients, which holds for $N\leq 5$.
It is now clear that a term in the sum of eq.~\eqref{eq:Wterm} 
transforms into the Hermitian conjugate of
%%%%%%%%%%%%%%%%%%%%%%%%%%
\begin{equation}
  g_s^{*} \left( Y_s(\tau) \psi_1 \ldots \psi_n \right)_{\mathbf{1},s} \, ,
\end{equation}
%%%%%%%%%%%%%%%%%%%
%
which should coincide with the original term due to the independence of
singlets in eq.~\eqref{eq:Wterm}. It now follows that
$g_s = g_s^{*}$, i.e., all coupling constants $g_s$ have to be real 
to conserve CP.

As a final remark, let us denote by $\tilde{g}_s$ the couplings written for a general 
basis and arbitrary normalisation of the modular form multiplets.
The CP constraint on $\tilde{g}_s$ is then more complicated, since 
the singlets of different bases coincide only up to normalisation 
factors, determined by the choice of normalisations of the Clebsch-Gordan 
coefficients and of the modular form multiplets.
Since the normalisation factors can differ between singlets, the 
corresponding couplings $\tilde{g}_s$ may require non-trivial phases 
to conserve CP.
These phases can be found directly by performing a basis transformation 
and matching $\tilde{g}_s$ to $g_s$ in the symmetric basis (and with proper modular form multiplet normalisation).

%=======================
\subsection{Implications of CP Invariance for the Mass Matrices}
%=======================

As a more concrete example, let us consider the Yukawa coupling term
%%%%%%%%%%%%%%%%%%%%%%%%
\begin{equation}
  W_{L} = \sum_s g_s \left( Y_s(\tau) E^c L H_d \right)_{\mathbf{1},s} \, ,
\end{equation}
%%%%%%%%%%%%%%%%
%
which gives rise to the charged lepton mass matrix.
Here $E^c$ is a modular symmetry multiplet of $SU(2)$ charged lepton singlets, $L$ is a modular symmetry multiplet of $SU(2)$ lepton doublets, and $H_d$ is a Higgs doublet which transforms trivially under modular symmetry and whose neutral component acquires a VEV $v_d = \langle H_d^0 \rangle$ after electroweak symmetry breaking.

Expanding the singlets, one gets
%%%%%%%%%%%%%%%%%%%%%%%%%%
\begin{equation}
  W_{L} \,=\, \sum_s g_s \lambda^s_{ij}(\tau) E^c_i L_j H_d 
  \,\equiv\,  \lambda_{ij}(\tau) E^c_i L_j H_d \, ,
\end{equation}
%%%%%%%%%%%%%%%%%%%%%%%%%
%
where entries of the matrices $\lambda^s_{ij}(\tau)$ are formed 
from components
of the corresponding modular multiplets $Y_s(\tau)$.
In a general basis, superfields transform under CP as
%%%%%%%%%%%%%%%%%%%%%%%%%%%%%%%
\begin{equation}
  E^c \,\xrightarrow{CP}\, X_R^{*} \, \overline{E^c}\,, \quad
  L   \,\xrightarrow{CP}\, X_L \, \overline{L}\,, \quad
  H_d \,\xrightarrow{CP}\, \eta_d \, \overline{H_d} \, ,
\end{equation}
%%%%%%%%%%%%%%%%%%
%
and we set $\eta_d = 1$ without loss of generality.
It follows that
%%%%%%%%%%%%%%%%%%%%%%%%
\begin{equation}
  W_{L} \,\xrightarrow{CP}\, \left( X_R^{\dagger} \, \lambda(-\tau^{*}) \, X_L \right)_{ij} \overline{E_i^c}\, \overline{L_j}\, \overline{H_d} \, ,
\end{equation}
%%%%%%%%%%%%%%%%%%%%%%%%%%%
%
so that CP conservation implies
%%%%%%%%%%%%%%%%%%%%%%%
\begin{equation}
  X_R^{\dagger} \, \lambda(-\tau^{*}) \, X_L = \lambda^{*} (\tau) \, .
\end{equation}
%%%%%%%%%%%%%%%%%%%%%%%%%%%%%%%%%

The resulting charged lepton mass matrix $M_e = v_d \lambda^{\dagger}$ (written in the left-right convention) satisfies
%%%%%%%%%%%%%%%%%%%%%%%%
\begin{equation}
  X_L^{\dagger} \, M_e(-\tau^{*}) \, X_R = M_e^*(\tau) \, ,
  \label{eq:Me_constraint}
\end{equation}
%%%%%%%%%%%%%%%%%%%%%%%%%%%
%
which coincides with the corresponding constraint in the case 
of CP invariance combined with discrete flavour symmetry, apart from the 
fact that now the mass matrix depends on the modulus $\tau$ which 
also transforms under CP.
Similarly, for the neutrino Majorana mass matrix $M_{\nu}$ one has
%%%%%%%%%%%%%%%%%%%%%%%%%%%%%%%%
\begin{equation}
  X_L^T M_{\nu}(-\tau^{*}) X_L = M_{\nu}^*(\tau) \, .
  \label{eq:Mnu_constraint}
\end{equation}
%%%%%%%%%%%%%%%%%%%%%%%%%%%%%%
%
Note that matrix $X_L$ is the same in eqs.~\eqref{eq:Me_constraint} and \eqref{eq:Mnu_constraint} since left-handed charged leptons $l_L$ and left-handed neutrinos $\nu_{lL}$ form an electroweak $SU(2)$ doublet $L$, so they transform uniformly both under CP and modular transformations:
%%%%%%%%%%%%%%%%%%%%%%%%%%
\begin{equation}
  X_{l_L} = X_{\nu_{lL}} \equiv X_L\,, \quad
  \rho_{l_L}(\gamma) = \rho_{\nu_{lL}}(\gamma) \equiv \rho_L(\gamma)\,, \quad
  k_{l_L} = k_{\nu_{lL}} \equiv k_L \, .
\end{equation}
%%%%%%%%%%%%%%%%%%%%%%
%
This can also be found directly from the form of the charged current (CC) weak interaction Lagrangian
%%%%%%%%%%%%%%%%%%%%%%%%%%%%%%%
\begin{equation}
  \mathcal{L}_{\text{CC}} = -\frac{g}{\sqrt{2}} \sum_{l=e,\mu,\tau} \overline{l_L} \, \gamma_{\alpha} \, \nu_{lL} \, W^{\alpha \dagger} + \text{h.c.}
\end{equation}
%%%%%%%%%%%%%%%%%%%%%
%
by ensuring its CP invariance.%
\footnote{Since the original superfields in a modular-invariant theory are not normalised canonically,
there is actually a prefactor of $(2 \im\tau)^{-k_L}$ in the CC weak interaction Lagrangian which originates from the Kähler potential.
This prefactor is necessary for modular invariance in order to compensate the weights $k_{l_L} = k_{\nu_{lL}} \equiv k_L$.}

In a symmetric basis 
$X_L = X_R = \mathbbm{1}$, 
the constraints on the mass matrices simplify to
%%%%%%%%%%%%%%%%%%%%%
\begin{equation}
  M_e(-\tau^{*}) = M_e^*(\tau), \quad
  M_{\nu}(-\tau^{*}) = M^*_{\nu}(\tau) \,,
  \label{eq:Me_Mnu_sym_basis_constraint}
\end{equation}
%%%%%%%%%%%%%%%%%%%%%%%%%%%
%
which further reduce to reality of the couplings. 
Namely, for the charged lepton mass matrix one has
%%%%%%%%%%%%%%%%%%%%%%%%%%
\begin{equation}
  \begin{aligned}
  M_e(-\tau^{*}) = v_d \sum_s g_s^{*} \, (\lambda^s)^{\dagger}(-\tau^{*})
  = v_d \sum_s g_s^{*} \, (\lambda^s)^T(\tau) \, ,\\
  M_e^*(\tau) = \left(v_d \sum_s g_s^{*} \, (\lambda^s)^{\dagger}(\tau) \right)^{*}
  = v_d \sum_s g_s \, (\lambda^s)^T(\tau) \, .
\end{aligned}
\end{equation}
%%%%%%%%%%%%%%%%%%%%%%%%%%%
%
Clearly, CP invariance requires $g_s = g_s^{*}$,
since $\lambda^s(\tau)$ are linearly independent matrices,
which in turn is guaranteed by independence of the singlets.

%=======================
\subsection{CP-Conserving Values of the Modulus \texorpdfstring{$\tau$}{τ}}
\label{sec:CPconsTau}
%=======================

In a CP-conserving modular-invariant theory both CP and modular symmetry are broken spontaneously by the VEV of the modulus $\tau$.
However, there exist certain values of $\tau$ which conserve CP, while breaking the modular symmetry.
Obviously, this is the case if $\tau$ is left invariant by CP, i.e.
%%%%%%%%%%%%%%%%%%%
\begin{equation}
  \tau \xrightarrow{CP} -\tau^{*} = \tau \, ,
  \label{eq:CP_invariant_tau}
\end{equation}
%%%%%%%%%%%%%%%%%%%%%%
%
meaning that $\tau$ lies on the imaginary axis, $\re\tau = 0$.
In a symmetric basis one then has
%%%%%%%%%%%%%%%%%%%%%%%%%%%%%
\begin{equation}
  M_e(\tau) = M_e^*(\tau)\,, \quad
  M_{\nu}(\tau) = M_{\nu}^*(\tau) \, ,
\end{equation}
%%%%%%%%%%%%%%%%%%%%%
%
as can be seen from eq.~\eqref{eq:Me_Mnu_sym_basis_constraint}.
The resulting mass matrices are real and the corresponding 
CPV phases are trivial,
such that $\sin\delta = \sin\alpha_{21} = \sin\alpha_{31} = 0$
in the standard parametrisation~\cite{PDG2019} of the PMNS mixing matrix.

Let us now consider a point $\gamma \tau$ in the plane of the modulus
related to a CP-invariant point $\tau = -\tau^{*}$ by a modular transformation $\gamma$.
This point is physically equivalent to $\tau$ due to modular invariance and therefore it should also be CP-conserving.
However, $\gamma \tau$ does not go to itself under CP.
Instead, one has
%%%%%%%%%%%%%%%%%%%%%%%%%%%
\begin{equation}
\gamma \tau \,\xrightarrow{CP}\,
(\gamma \tau)_{CP} \,=\, u(\gamma)\, \tau_{CP} \,=\, u(\gamma)\, \tau 
\,=\, u(\gamma)\gamma^{-1} \gamma \tau\,,
\end{equation}
%%%%%%%%%%%%%%%%%%%%%%%%%%%
%
so the resulting CP-transformed value $(\gamma \tau)_{CP}$
is related to the original value $\gamma \tau$ by a modular 
transformation $u(\gamma) \gamma^{-1}$.

Hence, it is natural to expect that a value of $\tau$ conserves CP if it is left invariant by CP \textit{up to a modular transformation}, i.e.,
%%%%%%%%%%%%%%%%%%%%%%%%%%%
\begin{equation}
  \tau \xrightarrow{CP} -\tau^{*} = \gamma \tau
  \label{eq:CP_conserving_tau}
\end{equation}
%%%%%%%%%%%%%%%%%%%%%%%%%%
%
for some $\gamma \in \overline{\Gamma}$.%
\footnote{
A similar condition has been derived in Ref.~\cite{Dent:2001cc} in the
context of string theories (in which the CP symmetry represents a
discrete gauge symmetry), postulating the action of CP on the
compactified directions.
}
Indeed, one can check that modular invariance of the mass terms requires the mass matrices to transform
under a modular transformation as
%%%%%%%%%%%%%%%%%%%%%%%%%%%%%
\begin{equation}
  \begin{aligned}
  M_e(\tau) & \,\xrightarrow{\gamma}\,
  M_e(\gamma \tau) = \rho_L(\gamma) \, M_e(\tau) \, \rho_E^T(\gamma) \, ,\\
  M_\nu(\tau) & \,\xrightarrow{\gamma}\,
  M_\nu(\gamma \tau) = \rho^{*}_L(\gamma) \, M_\nu(\tau) \, \rho_L^{\dagger}(\gamma) \, ,
\end{aligned}
\label{eq:Me_Mnu_modular_transform}
\end{equation}
%%%%%%%%%%%%%%%%%%%%%%%%%%%%%%
%
where $\rho_L$ and $\rho_E$ 
are the representation matrices for the $SU(2)$ lepton doublet $L$ and charged lepton singlets $E^c$, respectively.
We have also taken into account the rescaling of fields due to the non-canonical form
of the Kähler potential~\eqref{eq:Kahler}, which leads to cancellation of the modular weights in the transformed mass matrices.

It is clear from eq.~\eqref{eq:Me_Mnu_modular_transform} that mass eigenvalues are unaffected by the replacement $\tau \to \gamma \tau$ in the mass matrices.
Moreover, the unitary rotations $U_e$ and $U_{\nu}$ diagonalising the mass matrices $M_e M_e^{\dagger}$ and $M_{\nu}$ respectively transform as
%%%%%%%%%%%%%%%%%%%%%%%%%%
\begin{equation}
  U_e \xrightarrow{\gamma} \rho_L \, U_e, \quad
  U_{\nu} \xrightarrow{\gamma} \rho_L \, U_{\nu} \, ,
\end{equation}
%%%%%%%%%%%%%%%%%%%%%%%%
%
so the PMNS mixing matrix $U_{\text{PMNS}} = U_e^{\dagger} \, U_{\nu}$ 
does not change.
This means that the mass matrices evaluated at points $\tau$ and $\gamma \tau$ lead to the same values of physical observables.

If we now consider a value of $\tau$ which satisfies eq.~\eqref{eq:CP_conserving_tau}, then in a symmetric basis we have
%%%%%%%%%%%%%%%%%%%%%%%%%%
\begin{equation}
  M_e(\gamma \tau) = M_e^*(\tau)\,, \quad
  M_{\nu}(\gamma \tau) = M_{\nu}^*(\tau) \, ,
\end{equation}
%%%%%%%%%%%%%%%%%%%
%
from eq.~\eqref{eq:Me_Mnu_sym_basis_constraint}.
It follows from the above discussion that the observables 
evaluated at $\tau$ coincide with their complex conjugates,
hence CPV phases are trivial ($0$ or $\pi$).

To find all points satisfying eq.~\eqref{eq:CP_conserving_tau}, it is sufficient to restrict ourselves 
to the fundamental domain~$\mathcal{D}$ of the modular group given by
%%%%%%%%%%%%%%%%%%%%%%%
\begin{equation}
\mathcal{D} \, = \, 
\left\{ \tau \in \mathbb{C} :~~\im \tau > 0\,,~~|\re \tau| \leq \frac{1}{2}\,,~~|\tau| \geq 1 \right\},
\label{eq:fund_domain}
\end{equation}
%%%%%%%%%%%%%%%%%%%%%%%
%
since all other points are physically equivalent to 
the points from~$\mathcal{D}$.
All CP-conserving values of $\tau$ outside the fundamental domain are related to the CP-conserving values of $\tau$ within the fundamental domain by modular transformations.

The interior of $\mathcal{D}$, which we denote as $\text{int}(\mathcal{D})$,
maps to itself under CP.
Apart from that, no two points from $\text{int}(\mathcal{D})$ are related
by any non-trivial modular transformation.
Therefore, if $\tau \in \text{int}(\mathcal{D})$,
then eq.~\eqref{eq:CP_conserving_tau} reduces to
eq.~\eqref{eq:CP_invariant_tau} and we find again $\re \tau = 0$.
The remaining possibility is that $\tau$ lies on the boundary
of $\mathcal{D}$.
Then it is easy to show that it also satisfies
eq.~\eqref{eq:CP_conserving_tau}, but with a non-trivial $\gamma$.
Namely, for the left vertical line we have $\tau = -1/2 + i\,y \,\xrightarrow{CP}\, 1/2 + i\,y = T\tau$,
while for the arc we have $\tau = e^{i\varphi} \,\xrightarrow{CP}\, -e^{-i\varphi} = S\tau$.

We conclude that CP-conserving values of $\tau$ are the imaginary axis
and the boundary of the fundamental domain $\mathcal{D}$.
These values can also be obtained by algebraically solving eq.~\eqref{eq:CP_conserving_tau},
by noticing that it implies $u(\gamma) \gamma = I$
and using the requirement that $\gamma\tau$ belongs to the fundamental domain.

Thus, at the left and right vertical boundaries as well as at 
the arc boundary of the fundamental domain 
and at the imaginary axis within it,
i.e., at $\tau_{LB} = -1/2 + i\,y$, $\tau_{RB} = 1/2 + i\,y$, $y\geq \sqrt{3}/2$, 
at $\tau_{AB} = e^{i\varphi}$, $\varphi = [60^\circ,120^\circ]$,
and at $\tau_{IA} = i\,y$, $y\geq 1$,
we have a $\mathbb{Z}_2^{CP}$ symmetry. At the left and right cusps of 
the fundamental domain, $\tau_L =  -1/2 + i\,\sqrt{3}/2$ and 
$\tau_R = 1/2 + i\,\sqrt{3}/2$, and at $\tau_C = i$, 
the
symmetry is enhanced to
$\mathbb{Z}_2^{CP}\times \mathbb{Z}_3^{ST}$, 
$\mathbb{Z}_2^{CP}\times \mathbb{Z}_3^{TS}$ and 
$\mathbb{Z}_2^{CP}\times \mathbb{Z}_2^{S}$, respectively~\cite{Novichkov:2018ovf}, 
while at $\tau_T = i\infty$, the
symmetry in the case of 
$\Gamma_N$ is enhanced to
$\mathbb{Z}_2^{CP}\times \mathbb{Z}_N^{T}$~\cite{Novichkov:2018ovf,Novichkov:2018yse}.

%%%%%%%%%%%%%%%%%%%%%%%
\section{Example: CP-Invariant Modular \texorpdfstring{$S_4$}{S4} Models}
\label{sec:s4}
%%%%%%%%%%%%%%%%%%%%%%%

To illustrate the use of gCP invariance combined with modular symmetry for model building, we consider modular $\Gamma_4 \simeq S_4$ models of lepton masses and mixing, 
in which neutrino masses are generated via the type~I seesaw mechanism.
Such models have been extensively studied in Ref.~\cite{Novichkov:2018ovf} in the context of plain modular symmetry without gCP invariance.
Here we briefly summarise the construction of Ref.~\cite{Novichkov:2018ovf} and investigate additional constraints on the models imposed by CP invariance,
having the modulus $\tau$ as the only potential source of CPV.

Representations of the superfields under the modular group are chosen as follows:
\begin{itemize}
\item Higgs doublets $H_u$ and $H_d$ (with $\langle H_u^0 \rangle = v_u$) are $S_4$ trivial singlets of zero weight:
$\rho_u = \rho_d \sim \mathbf{1}$ and $k_u = k_d = 0$;
\item lepton $SU(2)$ doublets $L$ and neutral lepton gauge singlets $N^c$ form $S_4$ triplets ($\rho_L \sim \mathbf{3}$ or $\mathbf{3'}$, $\rho_N \sim \mathbf{3}$ or $\mathbf{3'}$) of weights $k_L$ and $k_N$, respectively;
\item charged lepton singlets $E_1^c$, $E_2^c$ and $E_3^c$ transform as $S_4$ singlets ($\rho_{1,2,3} \sim \mathbf{1}$ or $\mathbf{1'}$) of weights $k_{1,2,3}$. 
\end{itemize}
The relevant superpotential terms read
\begin{equation}
  W = \sum_{i=1}^3 \alpha_i \left( E_i^c L \, Y^{(k_{\alpha_i})} \right)_{\mathbf{1}} H_d
  + g \left( N^c L \, Y^{(k_g)} \right)_{\mathbf{1}} H_u
  + \Lambda \left( N^c N^c \, Y^{(k_{\Lambda})} \right)_{\mathbf{1}} \, ,
\end{equation}
where $Y^{(k)}$ denotes a modular multiplet of weight $k$ and level 4, and a sum over all independent singlets with the coefficients $\alpha_i = (\alpha_i, \, \alpha_i', \, \ldots)$, $g = (g, \, g', \, \ldots)$ and $\Lambda = (\Lambda, \, \Lambda', \, \ldots)$ is implied.
It has been found in Ref.~\cite{Novichkov:2018ovf} that the minimal (in terms of the total number of parameters) viable choice of modular weights and representations is
\begin{equation}
  \begin{aligned}
  & k_{\alpha_1} = 2, \quad
  k_{\alpha_2} = k_{\alpha_3} = 4, \quad
  k_g = 2, \quad
  k_{\Lambda} = 0, \\
  & \rho_L \sim \mathbf{3}, \quad
  \rho_1 \sim \mathbf{1'}, \quad
  \rho_2 \sim \mathbf{1}, \quad
  \rho_3 \sim \mathbf{1'}, \quad
  \rho_N \sim \mathbf{3} \text{ or } \mathbf{3'} \, ,
\end{aligned}
\end{equation}
which leads to the superpotential of the form
\begin{equation}
  \begin{aligned}
  W & = \alpha \left( E_1^c L \, Y_{\mathbf{3'}}^{(2)} \right)_{\mathbf{1}} H_d
  + \beta \left( E_2^c L \, Y_{\mathbf{3}}^{(4)} \right)_{\mathbf{1}} H_d
  + \gamma \left( E_3^c L \, Y_{\mathbf{3'}}^{(4)} \right)_{\mathbf{1}} H_d \\
  & + g \left( N^c L \, Y_{\mathbf{2}}^{(2)} \right)_{\mathbf{1}} H_u
  + g' \left( N^c L \, Y_{\mathbf{3'}}^{(2)} \right)_{\mathbf{1}} H_u
  + \Lambda \left( N^c N^c \right)_{\mathbf{1}} \, ,
\end{aligned}
\end{equation}
where the multiplets of modular forms $Y_\mathbf{2,3'}^{(2)}$ and $Y_\mathbf{3,3'}^{(4)}$ 
have been derived in Ref.~\cite{Penedo:2018nmg}.
Here no sums are implied, since each singlet is unique, and the coefficients
$(\alpha,\beta,\gamma) = (\alpha_1,\alpha_2,\alpha_3)$,
$g$ and $\Lambda$ are real without loss of generality,
as the corresponding phases can be absorbed into the fields $E_1^c$, $E_2^c$, $E_3^c$, $L$ and $N^c$, respectively. 
Therefore, the only complex parameter of the theory is $g'/g$.
If a symmetric basis is used and the modular form multiplets are properly normalised,
then CP is conserved whenever
\begin{equation}
  \im \left(g'/g\right) = 0 \, .
  \label{eq:S4_CP_condition}
\end{equation}

The basis used in Ref.~\cite{Novichkov:2018ovf} is not symmetric.
One can check that it can be related to the symmetric basis here considered by the following transformation matrices $U_{\mathbf{r}}$:
\begin{equation}
  \begin{aligned}
    U_{\mathbf{1}} &= U_{\mathbf{1'}} = 1, \qquad
    U_{\mathbf{2}} = \frac{1}{\sqrt{2}}
    \begin{pmatrix}
      1 & 1 \\
      -i & i
    \end{pmatrix}, \\
    U_{\mathbf{3}} &= U_{\mathbf{3'}} = \frac{1}{2\sqrt{3}}
    \begin{pmatrix}
      1 & 0 & 0 \\
      0 & -e^{-i \pi/4} & 0 \\
      0 & 0 & -e^{i \pi / 4}
    \end{pmatrix}
    \begin{pmatrix}
      2 & 2 & 2 \\
      -2 & 1+\sqrt{3} & 1-\sqrt{3} \\
      -2 & 1-\sqrt{3} & 1+\sqrt{3}
    \end{pmatrix} \, .
  \end{aligned}
\end{equation}
By direct comparison of the singlets $( N^c L \, Y_{\mathbf{2}}^{(2)})_{\mathbf{1}}$ and $( N^c L \, Y_{\mathbf{3'}}^{(2)})_{\mathbf{1}}$ written in different bases, and taking into account an extra factor of $i$ arising from the 
normalisation of the modular form multiplets used in Ref.~\cite{Novichkov:2018ovf}, we find that
also in this basis CP invariance results in the condition~\eqref{eq:S4_CP_condition}.
In what follows, we report the parameter values in the basis of Ref.~\cite{Novichkov:2018ovf} for ease of comparison.

Through numerical search, five viable pairs of regions of the parameter space have been found in Ref.~\cite{Novichkov:2018ovf}, denoted as A and A$^{*}$, B and B$^{*}$, etc. with the starred regions corresponding to CP-conjugated models $\tau \to -\tau^{*}$, $(g'/g) \to (g'/g)^{*}$ predicting the opposite values of the CPV phases. 
Among these five pairs of regions only one pair (A and A$^{*}$, for which $\rho_N \sim \mathbf{3'}$) is consistent with the condition~\eqref{eq:S4_CP_condition}, and only a small portion of the parameter space is allowed.
We report the corresponding best fit values and the confidence intervals of the parameters and observables in Table~\ref{tab:S4_predictions}.
%%%%%%%%%%%%%%%%%%%%%%%%%%%%%%
\begin{table}
\centering
\renewcommand{\arraystretch}{1.2}
\begin{tabular}{c|ccc}
  \toprule
  & Best fit value & $2\sigma$ range & $3\sigma$ range \\
  \midrule
  $\re \tau$ & $\pm 0.09922$ & $\pm (0.0961 - 0.1027)$ & $\pm (0.09371 - 0.1049)$ \\
  $\im \tau$ & 1.016 & $1.015 - 1.017$ & $1.014 - 1.018$ \\
  $\beta/\alpha$ & 9.348 & $8.426 - 11.02$ & $7.845 - 12.25$ \\
  $\gamma/\alpha$ & 0.002203 & $0.002046 - 0.00236$ & $0.001954 - 0.00246$ \\
  $g'/g$ & $-0.02093$ & $-(0.01846 - 0.02363)$ & $-(0.01682 - 0.02528)$ \\
  $v_d\,\alpha$ [MeV] & 53.61 \\
  $v_u^2\, g^2 / \Lambda$ [eV] & 0.0135 \\
  \midrule
  $m_e/m_{\mu}$ & 0.004796 & $0.004454 - 0.005135$ & $0.004251 - 0.005351$ \\
  $m_{\mu} / m_{\tau}$ & 0.05756 & $0.0488 - 0.06388$ & $0.04399 - 0.06861$ \\
  $r$ & 0.02981 & $0.02856 - 0.0312$ & $0.02769 - 0.03212$ \\
  $\delta m^2$ [$10^{-5} \text{ eV}^2$] & 7.326 & $7.109 - 7.551$ & $6.953 - 7.694$ \\
  $|\Delta m^2|$ [$10^{-3} \text{ eV}^2$] & 2.457 & $2.421 - 2.489$ & $2.396 - 2.511$ \\
  $\sin^2 \theta_{12}$ & 0.305 & $0.2825 - 0.3281$ & $0.2687 - 0.3427$ \\
  $\sin^2 \theta_{13}$ & 0.02136 & $0.02012 - 0.02282$ & $0.0192 - 0.02372$ \\
  $\sin^2 \theta_{23}$ & 0.4862 & $0.4848 - 0.4873$ & $0.484 - 0.4882$ \\
  \midrule
  Ordering & NO \\
  $m_1$ [eV] & 0.01211 & $0.01195 - 0.01226$ & $0.01185 - 0.01236$ \\
  $m_2$ [eV] & 0.01483 & $0.01477 - 0.01489$ & $0.01473 - 0.01493$ \\
  $m_3$ [eV] & 0.05139 & $0.051 - 0.05172$ & $0.05074 - 0.05195$ \\
  $\textstyle \sum_i m_i$ [eV] & 0.07833 & $0.07774 - 0.07886$ & $0.07734 - 0.07921$ \\
  $|\langle m\rangle|$ [eV] & 0.01201 & $0.01187 - 0.01213$ & $0.01178 - 0.01221$ \\
  $\delta/\pi$ & $\pm 1.641$ & $\pm (1.633 - 1.651)$ & $\pm (1.627 - 1.656)$ \\
  $\alpha_{21}/\pi$ & $\pm 0.3464$ & $\pm (0.3335 - 0.3618)$ & $\pm (0.324 - 0.3713)$ \\
  $\alpha_{31}/\pi$ & $\pm 1.254$ & $\pm (1.238 - 1.271)$ & $\pm (1.229 - 1.283)$ \\
  \midrule
  $N \sigma$ & 1.012 \\
  \bottomrule
\end{tabular}
\caption{
  Best fit values along with $2\sigma$ and $3\sigma$ ranges of the parameters and observables
in the minimal CP-invariant modular $S_4$ model.
  CP symmetry is spontaneously broken by the VEV of the modulus $\tau$.} \label{tab:S4_predictions}
\end{table}
%%%%%%%%%%%%%%%%%%%%%%%%%%%%%%

%
%%%%%%%%%%%%%%%%
\begin{figure}[p]
  \centering
  \includegraphics[width=\textwidth]{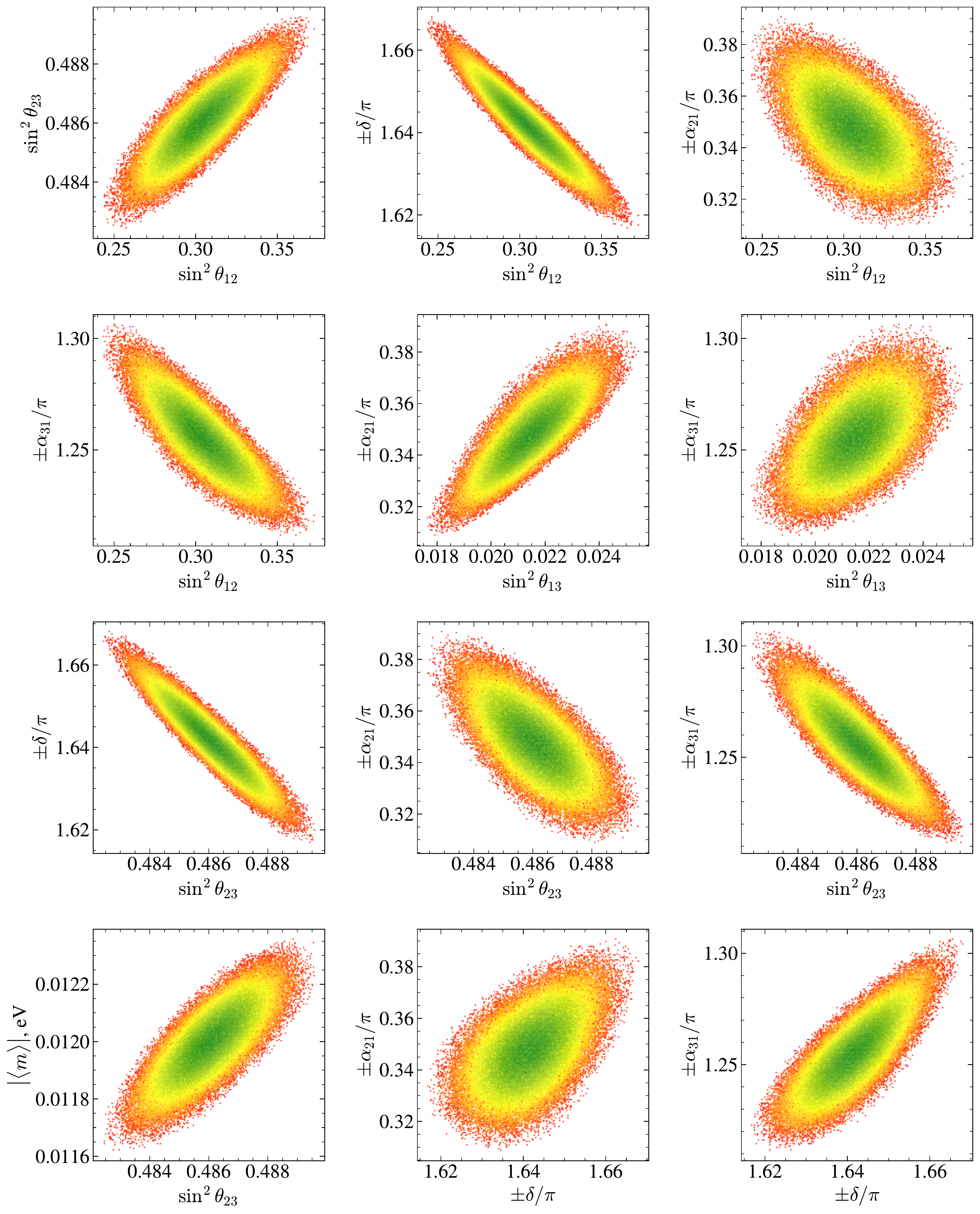}
  \caption{
    Correlations between pairs of observables in the minimal CP-invariant modular $S_4$ model.}
  \label{fig:correlations}
\end{figure}
%%%%%%%%%%%%%%%%
%
%
%%%%%%%%%%%%%%%%
\begin{figure}[p]
  \centering
  \includegraphics[width=\textwidth]{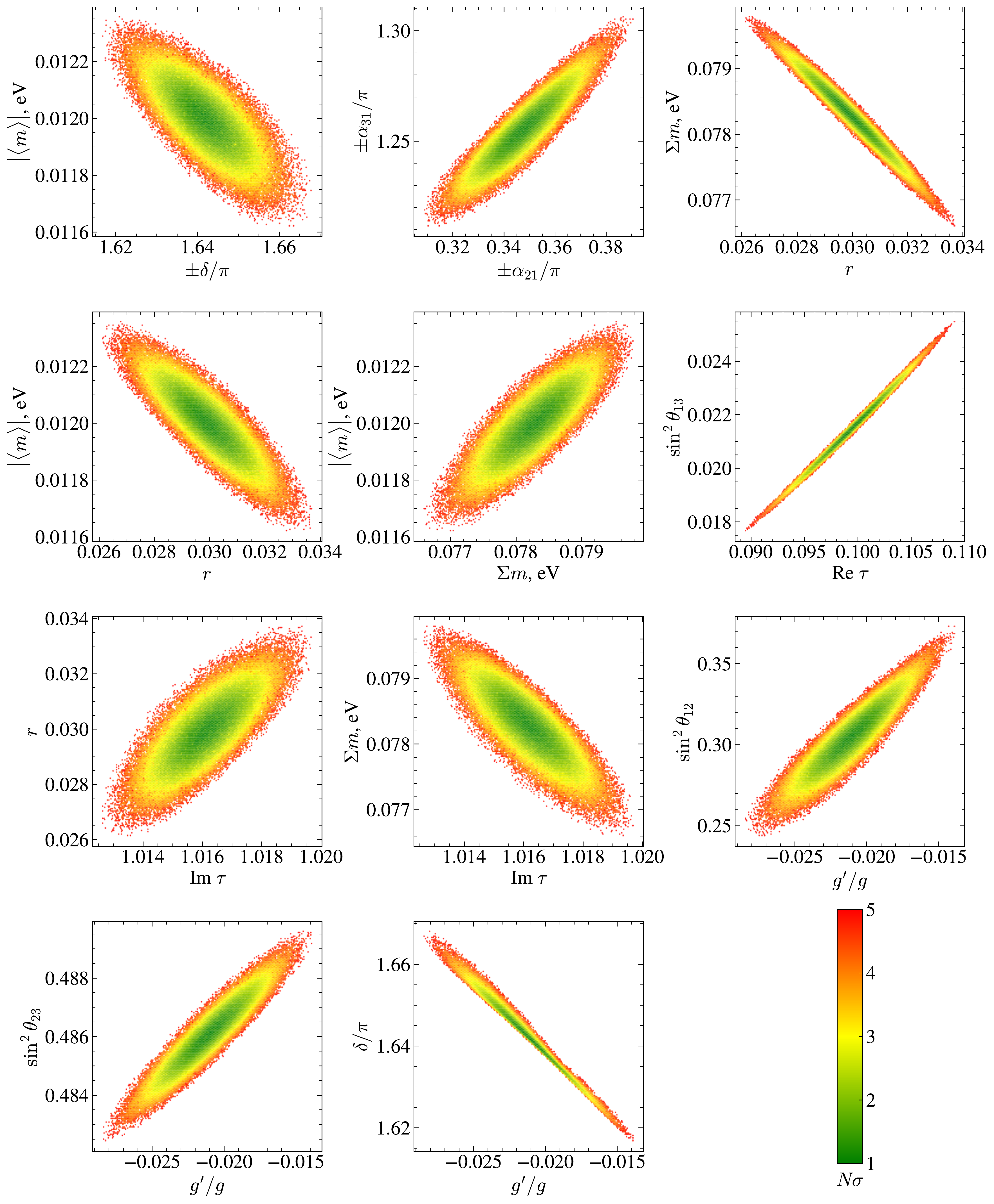}
  \caption{
    Correlations between pairs of observables (continued from Fig.~\ref{fig:correlations}) and between observables and parameters in the minimal CP-invariant modular $S_4$ model.}
  \label{fig:correlations2}
\end{figure}
%%%%%%%%%%%%%%%%
%

This minimal CP-invariant model, predicting 12 observables,
is characterised by 7 parameters: the 6 real parameters $v_d \, \alpha$, $\beta/\alpha$, $\gamma/\alpha$, $v_u^2 \, g^2 / \Lambda$, $g'/g$, $\im \tau$ and the phase $\re \tau$.%
\footnote{$\re \tau$ should be treated as a
phase since the dependence 
of Yukawa couplings and fermion mass matrices
on $\tau$ arises through powers of 
$\exp (2\pi i \tau / N) = \exp (\pi i \tau / 2)$.}
The three real parameters $v_d \, \alpha$, $\beta/\alpha$ and $\gamma/\alpha$ are fixed by fitting the three charged lepton masses.
The remaining three real parameters $v_u^2 \, g^2 / \Lambda$, $g'/g$, $\im \tau$ and the phase $\re \tau$ describe the nine neutrino observables: three neutrino masses, three neutrino mixing angles and three CPV phases.

As a result, this model has more predictive power than the original
model from Ref.~\cite{Novichkov:2018ovf}, which is described by the same parameters and an additional phase $\arg (g'/g)$.
In fact, the correlations between $\sin^2 \theta_{23}$, the neutrino masses and the CPV phases,
which were present in the original
model, now reduce to accurate predictions
of these observables at a few percent level.
This can be seen by comparing the ranges from Table~\ref{tab:S4_predictions} of the present article with Table~5a and Fig.~2 of Ref.~\cite{Novichkov:2018ovf}.
Apart from that, many correlations between pairs of observables and between observables and parameters arise. 
We report these correlations in Figs.~\ref{fig:correlations} and~\ref{fig:correlations2}.

\begin{figure}[t]
  \centering
  \includegraphics[width=\textwidth]{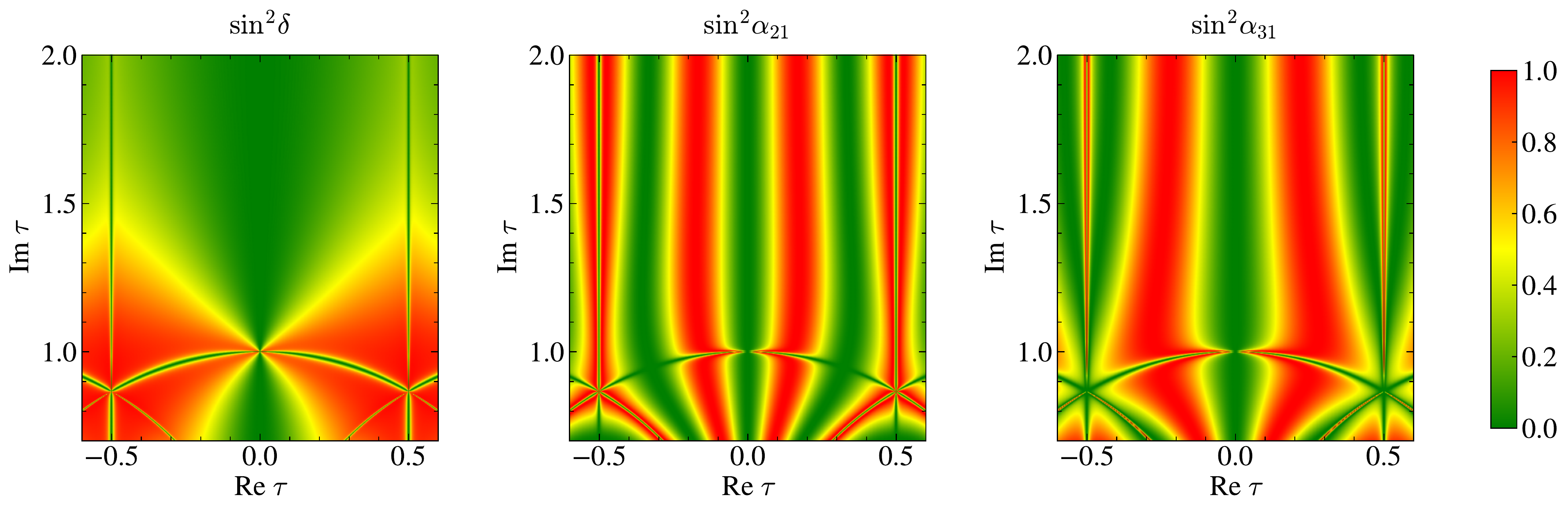}
  \caption{CP violation strength, measured as sine-squared of the CPV phases, for different values of the modulus $\tau$.
The boundary of the fundamental domain $\mathcal{D}$ defined in eq.~\eqref{eq:fund_domain} 
and the imaginary axis $\re\tau = 0$ conserve CP.  
  The ranges of $\re \tau$ and $\im \tau$ are chosen to extend slightly beyond the fundamental domain $\mathcal{D}$
  to make its boundary clearly visible.
}
  \label{fig:cpv_heatmaps}
\end{figure}

We also check numerically that CP invariance is restored for the CP-conserving values of $\tau$ derived in subsection~\ref{sec:CPconsTau}.
To achieve this, we vary the value of $\tau$ while keeping all other parameters fixed to their best fit values, and
present the resulting 
$\sin^2 \delta(\tau)$, $\sin^2 \alpha_{21}(\tau)$ and $\sin^2 \alpha_{31}(\tau)$ as heatmap plots in the $\tau$ plane 
in Fig.~\ref{fig:cpv_heatmaps}.
Notice that this variation is done for illustrative purposes only,
as it spoils the values of the remaining observables.
Those are in agreement with experimental data
only in a small region of the $\tau$ plane~\cite{Novichkov:2018ovf}.
The sine-squared of a phase measures the strength of CPV, with the value of 0 (shown with green colour) corresponding to no CPV 
and the value of 1 (shown with red colour) corresponding to maximal CPV. 
As anticipated, both the boundary of $\mathcal{D}$ and the imaginary axis conserve CP, appearing in green colour in Fig.~\ref{fig:cpv_heatmaps}.
However, even a small 
departure from a CP-conserving value of $\tau$ 
can lead to large CPV 
due to strong dependence of the observables on $\tau$.
This is noticeably the case in the vicinity of the boundary of the fundamental domain.
%
%%%%%%%%%%%%%%%%
%%%%%%%%%%%%%%%%
%

%%%%%%%%%%%%%%%%%%%%%%%%%%%%%%%%%
\section{Summary and Conclusions}
\label{sec:conclusions}
%%%%%%%%%%%%%%%%%%%%%%%%%%%%%%%%%

In the present article we have developed the formalism 
of combining modular and generalised CP (gCP) symmetries 
for theories of flavour. To this end the  
corresponding consistency conditions for the 
two symmetry transformations acting on the modulus $\tau$ and on the matter 
fields were derived. We have shown that these
consistency conditions  
imply that under the CP transformation 
the modulus $\tau \rightarrow n -\tau^*$ with integer $n$, and one can choose $n=0$ without loss of generality.
This transformation extends the modular group 
$\overline\Gamma \simeq PSL(2,\mathbb{Z})$ to 
$\overline\Gamma^* \simeq \overline\Gamma \rtimes \mathbb{Z}_2^{CP} 
\simeq PGL(2,\mathbb{Z})$. 
Considering the cases of the finite modular groups 
$\Gamma_N$ with $N=2,3,4,5$, which are isomorphic 
to the non-Abelian discrete groups $S_3$, $A_4$, $S_4$ and $A_5$, respectively, 
we have demonstrated that the gCP transformation matrix 
$X_\mathbf{r}$ realising a rotation in flavour space when acting on a multiplet $\psi(x)$ 
as $\psi(x) \to X_\mathbf{r} \,\overline\psi(x_{P})$, where $x_{P} = (t,-\mathbf{x})$, 
can always be chosen to be the identity matrix $\mathbbm{1}_\mathbf{r}$.
Assuming this choice and a proper normalisation of multiplets of modular forms $Y(\tau)$,
transforming in irreducible representations of the groups $\Gamma_N$ with $N=2,3,4,5$, we have shown that under CP these multiplets get complex conjugated.
As a consequence, we have found that gCP invariance implies that 
the constants $g$, which accompany each invariant singlet 
in the superpotential, must be real. 
Thus, the number of free parameters 
in modular-invariant models which also enjoy a gCP symmetry 
gets reduced, leading to a higher predictive power of such models. 
In these models, the only source of both modular symmetry breaking and CP violation is the VEV of the modulus $\tau$. 
We have further demonstrated that CP is conserved for the values of the modulus at the boundary of 
the fundamental domain and on the imaginary axis.

Finally, via the example of a modular $S_4$ model of lepton flavour with type I seesaw mechanism of neutrino mass generation,
we have illustrated the results obtained 
in the present study regarding the implications of gCP symmetry in modular-invariant theories of flavour.
This model was considered in Ref.~\cite{Novichkov:2018ovf} without the requirement of gCP invariance. 
We have shown that imposing the latter leads to much reduced 
ranges of allowed values of the neutrino mass and mixing parameters as well as to much stronger correlations between 
the different neutrino-related observables (Table~\ref{tab:S4_predictions} and Figs.~\ref{fig:correlations} and \ref{fig:correlations2}).

%%%%%%%%%%%%%%%%%%%%%%%
\section*{Acknowledgements}
%%%%%%%%%%%%%%%%%%%%%%%
P.P.N. would like to thank Kavli IPMU, where part of this work was 
carried out, for its hospitality.
This project has received funding from the European Union's Horizon 2020 
research and innovation programme under the Marie 
Skłodowska-Curie grant agreements No 674896 
(ITN Elusives) and No 690575 (RISE InvisiblesPlus).
This work was supported in part 
by the INFN program on Theoretical Astroparticle Physics (P.P.N. and S.T.P.)
and by the  World Premier International Research Center
Initiative (WPI Initiative, MEXT), Japan (S.T.P.).
The work of J.T.P.~was supported by
Fundação para a Ciência e a Tecnologia (FCT, Portugal) through the projects
CFTP-FCT Unit 777 (UID/FIS/00777/2013 and UID/FIS/00777/2019) 
and PTDC/FIS-PAR/29436/2017 
which are partially funded through POCTI (FEDER), COMPETE, QREN and EU.

\pagebreak

%%%%%%%%%%%%%%%%%%%%%%%
\appendix
%%%%%%%%%%%%%%%%%%%%%%%

%%%%%%%%%%%%%%%%%%%%%%%%%%%%%%%%%
\section{Spinors and Superfields under CP}
\label{app:SUSYCP}
%%%%%%%%%%%%%%%%%%%%%%%%%%%%%%%%%

For clarity, in this Appendix alone chiral superfields will be denoted with a tilde to distinguish them from scalar and fermion fields.
Under CP, a 4-component spin-$1/2$ flavour multiplet $\Psi$ transforms as
%%%%%%%%%%%%%%%%%%%%%%%%%
\begin{align}
\Psi_i(x) \,\xrightarrow{CP}\, i\,\left(X_{\Psi}\right)_{ij}\, \gamma^0 \, C\, \overline{\Psi_j}^T(x_P)\,,
\label{eq:Dirac_CP}
\end{align}
%%%%%%%%%%%%%%%%%%%%%%%%%%
%
where $x = (t,\mathbf{x})$, $x_P = (t,-\mathbf{x})$,
$X_{\Psi}$ is a unitary matrix in flavour space and $C$ is
the charge conjugation matrix, 
satisfying $C^{-1} \gamma_\mu C = -\,\gamma^T_\mu$, 
which implies $C^T = - C$. 
These two properties of 
$C$ do not depend on the representation of the
$\gamma$-matrices.
We can further consider $C$ to be unitary, $C^{-1} = C^\dagger$, 
without loss of generality.
The factor of $i$ in eq.~\eqref{eq:Dirac_CP} is a convention employed 
consistently throughout this Appendix.
Spacetime coordinates are henceforth omitted.
For the two-component formalism widely used in SUSY we are going to discuss  
in the present Appendix it proves convenient to consider
the Weyl basis for $\gamma^\mu$ matrices.
In this basis, the matrix $C$ 
is real, so that $C = - C^\dagger = - C^{-1}$.

One may write a 4-component spinor $\Psi$ in terms of two
Weyl 2-spinors,
%%%%%%%%%%%%%%%%%%%%%%%%%%%%%%%
\begin{align}
\Psi = \begin{pmatrix}
\psi_{\alpha} \\[2mm] \overline \phi^{\dot \alpha}
\end{pmatrix},
\qquad
\Psi^c \equiv C\,\overline{\Psi}^T = \begin{pmatrix}
\phi_{\alpha} \\[2mm] \overline{\psi}^{\dot \alpha}
\end{pmatrix},
\label{eq:DiracWeyl}
\end{align}
%%%%%%%%%%%%%%%%%%
%
with dotted and undotted indices shown explicitly.
Bars on 2-spinors denote conjugation,
e.g.~$\overline{\psi}^{\dot\alpha} = {\delta^{\dot \alpha}}_\alpha (\psi^\alpha)^*$.
Notice that for 4-spinors
$\overline\Psi \equiv \Psi^\dagger A$, with $A$ numerically 
equal to $\gamma^0$ but with a different index structure. One has:
%%%%%%%%%%%%%%%%%%%%%%%%%%%%
\begin{align}
A &= \begin{pmatrix}
0 & {\delta^{\alpha}}_{\dot \beta} \\
{\delta_{\dot \alpha}}^{\beta} & 0
\end{pmatrix},
\qquad
C = \begin{pmatrix}
\epsilon_{\alpha \beta} & 0 \\
0 & \epsilon^{\dot\alpha \dot\beta} 
\end{pmatrix},
\quad\text{with }
\epsilon^{\dot\alpha\dot\beta} = -\epsilon_{\alpha\beta}= \begin{pmatrix} 0 & 1 \\ -1 & 0 \end{pmatrix},
\\[1mm]
\gamma^\mu &= 
\begin{pmatrix}
0 & {(\sigma^\mu)}_{\alpha \dot \beta} \\
(\overline \sigma^\mu)^{\dot \alpha \beta} & 0
\end{pmatrix} 
\quad\Rightarrow\quad
\gamma^0 = \begin{pmatrix}
0 & {(\sigma^0)}_{\alpha \dot \beta} \\
{(\overline\sigma^0)}^{\dot \alpha \beta} & 0 
\end{pmatrix}.
\end{align}
%
%%%%%%%%%%%%%%%%%%%%%%%%%%%%%%%%%%%%
%
Disregarding the spinor-index structure, one has
$C = i \gamma^0 \gamma^2$,%
\footnote{The matrix $C = i \gamma^0 \gamma^2$ has the same form in the usual Dirac representation of $\gamma$-matrices.}
$\gamma^5 \equiv i \gamma^0\gamma^1\gamma^2\gamma^3 = \diag(-\mathbbm{1},\mathbbm{1})$,
$\sigma^\mu = (\sigma^0, \sigma^i)= (\sigma_0, -\sigma_i)$ and $\overline\sigma^\mu = (\sigma_0, \sigma_i)$,
with $\sigma_0 = \mathbbm{1}$ and the $\sigma_i$ being the Pauli matrices.
The chiral projection operators are defined as 
$P_{L,R} = (\mathbbm{1} \mp \gamma^5)/2$.

From the CP transformation \eqref{eq:Dirac_CP} of a 4-spinor $\Psi$,
one can obtain the CP transformation of the Weyl spinors $\psi$ and $\phi$ in eq.~\eqref{eq:DiracWeyl}:
\begin{align}
\psi_{i\alpha} \,\xrightarrow{CP}\, i\,\left(X_{\Psi}\right)_{ij}\, {(\sigma^0)}_{\alpha \dot \beta} \,\overline\psi_j^{\dot \beta} 
\qquad &\Rightarrow \qquad
\overline\psi_i^{\dot\alpha} \,\xrightarrow{CP}\, i\,\left(X_{\Psi}\right)^*_{ij}\, {(\overline\sigma^0)}^{\dot \alpha \beta} \,\psi_{j \beta} 
\,,
\label{eq:Weyl_CP}\\
\overline\phi_i^{\dot\alpha} \,\xrightarrow{CP}\, i\,\left(X_{\Psi}\right)_{ij}\, {(\overline\sigma^0)}^{\dot \alpha \beta} \,\phi_{j \beta} 
\qquad &\Rightarrow \qquad
\phi_{i\alpha} \,\xrightarrow{CP}\, i\,\left(X_{\Psi}\right)^*_{ij}\, {(\sigma^0)}_{\alpha \dot \beta} \,\overline\phi_j^{\dot \beta} \,.
\end{align}

For the (chiral) fields in the lepton sector, in particular, one has:
\begin{align}
    \ell_{iL} \equiv \begin{pmatrix} L_{i \alpha}\\ 0\end{pmatrix}
    \,\xrightarrow{CP}\, i\,\left(X_L\right)_{ij}\, \gamma^0 \, C\, \overline{\ell_{jL}}^T
    \quad&\Rightarrow \quad 
    L_{i \alpha}\,\xrightarrow{CP}\, i \, \left(X_L\right)_{ij} \, (\sigma^0)_{\alpha\dot{\beta}} \,\overline{L_j}^{\dot{\beta}}\,,
\label{eq:lepton_spinors_left}
\\[1mm]
    e_{iR} \equiv \begin{pmatrix} 0\\ \overline{E^c_i}^{\dot\alpha}  \end{pmatrix}
    \,\xrightarrow{CP}\, i\,\left(X_R\right)_{ij}\, \gamma^0 \, C\, \overline{e_{jR}}^T
    \quad&\Rightarrow \quad 
    {E^c_i}_{\alpha}\,\xrightarrow{CP}\, i\, \left(X_R\right)^*_{ij}\, (\sigma^0)_{\alpha\dot{\beta}}\, \overline{E^c_j}^{\dot{\beta}}\,,
\\[1mm]
    \nu_{iR} \equiv \begin{pmatrix} 0\\ \overline{N^c_i}^{\dot\alpha}  \end{pmatrix}
    \,\xrightarrow{CP}\, i\,\left(X_N\right)_{ij}\, \gamma^0 \, C\, \overline{\nu_{jR}}^T
    \quad&\Rightarrow \quad 
    {N^c_i}_{\alpha}\,\xrightarrow{CP}\, i\, \left(X_N\right)^*_{ij}\, (\sigma^0)_{\alpha\dot{\beta}}\, \overline{N^c_j}^{\dot{\beta}}\,,
\label{eq:lepton_spinors_right}
\end{align}
where $\ell_{iL}$, $e_{iR}$ and $\nu_{iR}$ ($L_i$, $E^c_i$ and $N^c_i$) denote
lepton doublet, charged lepton singlet and neutrino singlet 4(2)-spinors, respectively.
It is then straightforward to find the transformation of a pair of contracted spinors, e.g.:
\begin{equation}
\begin{aligned}
    E^c_i L_j \,=\, \epsilon^{\alpha \beta} E^c_{i\beta} L_{j \alpha}
    \,\,\xrightarrow{CP}\,\, &\left(X_R\right)^*_{ik} \left(X_L\right)_{jl}
    \left[- (\sigma^{0T})_{\dot \beta \alpha} \epsilon^{\alpha \beta} (\sigma^0)_{\beta \dot \alpha} \right] \overline{E^c_k}^{\dot{\alpha}} \overline{L_l}^{\dot{\beta}}
    \\ =\, &\left(X_R\right)^*_{ik} \left(X_L\right)_{jl}
    \overline{E^c_k}\,\overline{L_l}\,.
\label{eq:contract}
\end{aligned}
\end{equation}

In the framework of rigid $\mathcal{N} = 1$ SUSY in 4 dimensions, 
taking the Graßmann coordinates $\theta_{\alpha}$ and $\overline{\theta}^{\dot\alpha}$
to transform under CP as all other Weyl spinors, i.e.
\begin{align}
\theta_{\alpha} \,\xrightarrow{CP}\, i\, {(\sigma^0)}_{\alpha \dot \beta} \,\overline\theta^{\dot \beta} 
\quad\Rightarrow \quad 
\overline\theta^{\dot\alpha} \,\xrightarrow{CP}\, i\, {(\overline\sigma^0)}^{\dot \alpha \beta} \,\theta_{\beta} \,,
\label{eq:grassmann_CP}
\end{align}
one can obtain a consistent CP transformation of a chiral superfield $\widetilde \psi = \varphi + \sqrt{2}\, \theta \psi - \theta^2 F$, with
$F$ being an auxiliary field and $\varphi$ denoting the scalar lowest component of the superfield, expected to transform under CP in the same way as the latter.
Indeed, using eqs.~\eqref{eq:grassmann_CP} and \eqref{eq:Weyl_CP}, one sees that consistency implies
\begin{align}
\widetilde \psi_i \,\xrightarrow{CP}\, \left(X_{\Psi}\right)_{ij}\,\overline{\widetilde{\!\psi}}_j\,,
\label{eq:superfield_CP}
\end{align}
with $\varphi_i \xrightarrow{CP} \left(X_{\Psi}\right)_{ij} \varphi_j^*$ and $F_i \xrightarrow{CP} \left(X_{\Psi}\right)_{ij} F_j^*$,
since  $\theta^2\xrightarrow{CP} \overline\theta^2$, and where the bar denotes the Hermitian conjugated superfield.
We thus see that the superpotential, up to unitary rotations in flavour space,
is exchanged under CP with its conjugate.%
\footnote{One can check that the Graßmann integration $\int d^2 \theta$ is exchanged with its conjugate $\int d^2 \overline\theta$.}

Given the CP transformations of spinors in eqs.~\eqref{eq:lepton_spinors_left}\,--\,\eqref{eq:lepton_spinors_right}, 
one then has, for the chiral superfields $\widetilde L_i$, $\widetilde E^c_i$ and $\widetilde N^c_i$ in the lepton sector:
\begin{alignat}{4}
&\widetilde L_i &&\,\xrightarrow{CP}\, \left(X_L\right)_{ij}\,\overline{\widetilde{\!L}}_j
    \qquad&&\Rightarrow \qquad 
\overline{\widetilde{\!L}}_i &&\xrightarrow{CP}\, \left(X_L\right)^*_{ij}\,\widetilde L_j\,,
\\[1mm]
&\widetilde E^c_i &&\,\xrightarrow{CP}\, \left(X_R\right)^*_{ij}\, \overline{\!\widetilde{\!E^c_j\!}\!}
    \qquad&&\Rightarrow \qquad 
\overline{\!\widetilde{\!E^c_i\!}\!} &&\xrightarrow{CP}\, \left(X_R\right)_{ij}\,\widetilde E^c_j\,,
\\[1mm]
&\widetilde N^c_i &&\,\xrightarrow{CP}\, \left(X_N\right)^*_{ij}\, \overline{\!\widetilde{\!N^c_j\!}\!}
    \qquad&&\Rightarrow \qquad 
\overline{\!\widetilde{\!N^c_i\!}\!} &&\xrightarrow{CP}\, \left(X_N\right)_{ij}\,\widetilde N^c_j\,.
\end{alignat}

\vfill\pagebreak
%%%%%%%%%%%%%%%%%%%%%%%
\section{Group Theory of \texorpdfstring{$\Gamma_{N\leq 5}$}{ΓN≤5}}
\label{app:group}
%%%%%%%%%%%%%%%%%%%%%%%
%=======================
\subsection{Order and Irreducible Representations}
\label{app:orderirreps}
%=======================
The finite modular groups $\Gamma_N = \overline\Gamma / \overline\Gamma(N) \simeq PSL(2,\mathbb{Z}_N)$,
with $N > 1$, can be defined by two generators $S$ and $T$ satisfying the relations:
\begin{align}
S^2 = T^N = (ST)^3 = I\,.
\end{align}
The order of these groups is given by  (see, e.g., \cite{Schoeneberg:1974md}):
\begin{equation}
  |\Gamma_N| = 
\begin{cases}
6 & \text{for } N = 2\,,\\[2mm]
\displaystyle\frac{N^3}{2} \prod_{p|N} \left(1-\frac{1}{p^2}\right)  & \text{for } N > 2\,,
\end{cases}  
\end{equation}
where the product is over prime divisors $p$ of $N$.

It is straightforward to show that $|\Gamma_N|$ is even for all $N$. Decomposing $N>2$ in its unique prime factorisation,
$N = \prod_{i=1}^k p_i^{n_i}$ with $n_i \geq 1$, one has:
\begin{equation}
2 \, |\Gamma_{N > 2}| = N \,\prod_{i = 1}^k p_i^{2(n_i-1)}\left(p_i^2-1\right)\,.
\end{equation}
The group order will be even if $2 |\Gamma_{N > 2}|$ is a multiple of 4.
This is trivially verified when $N$ is a power of 2. If $N$ is not a power of 2,
then at least one of its prime factors is odd, say $p_j$. Since $(n^2 -1) \equiv 0\, (\text{mod } 4)$ for odd $n$,
it follows that the group order is even also in this case, as $(p_j^2 -1)$ divides $2|\Gamma_{N > 2}|$.

In what follows, we focus on the case $N \leq 5$. The orders and irreducible representations of these groups are listed in Table~\ref{tab:ord_irreps}.
\begin{table}
\centering
\renewcommand{\arraystretch}{1.2}
\begin{tabular}{lcccc}
  \toprule 
  $\Gamma_N\qquad$ & $\Gamma_2 \simeq S_3$ & $\Gamma_3 \simeq A_4$ & $\Gamma_4 \simeq S_4$ & $\Gamma_5 \simeq A_5$ \\
  \midrule
  $|\Gamma_N|$ & 6 & 12 & 24 & 60 \\
  irreps  & $\mathbf{1}, \mathbf{1'}, \mathbf{2}$ & $\mathbf{1}, \mathbf{1'}, \mathbf{1''}, \mathbf{3}$ & $\mathbf{1}, \mathbf{1'}, \mathbf{2}, \mathbf{3}, \mathbf{3'}$ & $\mathbf{1}, \mathbf{3}, \mathbf{3'}, \mathbf{4}, \mathbf{5}$ \\
  \bottomrule
\end{tabular}
\caption{Orders and irreducible representations for finite modular groups $\Gamma_{N}$ with $N\leq 5$.}
\label{tab:ord_irreps}
\end{table}

%=======================
\subsection{Symmetric Basis for Group Generators}
\label{app:sym_basis}
%=======================
For the groups $S_3$, $A_4$, $S_4$ and $A_5$,
we explicitly give below the basis for representation matrices in which the group generators $S$ and $T$ are represented by symmetric matrices,
see eq.~\eqref{eq:sym_generators}, for all irreducible representations $\mathbf{r}$ (see, e.g., Refs.~\cite{Ishimori:2010au,Feruglio:2017spp,Altarelli:2009gn,Ding:2011cm}).
\subsubsection{\texorpdfstring{$\Gamma_2 \simeq S_3$}{Γ2=S3}}
\begin{align}
\mathbf{1}:&\quad \rho(S)= 1 ,\quad \rho(T)= 1\,, \\
\mathbf{1'}:&\quad \rho(S)= -1,\quad \rho(T)= -1\,,\\
\mathbf{2}:&\quad \rho(S)= 
\frac{1}{2}
\begin{pmatrix}
-1 & -\sqrt{3} \\ -\sqrt{3} & 1
\end{pmatrix}
,\quad \rho(T)= 
\begin{pmatrix}
1 & 0 \\ 0 & -1 
\end{pmatrix}\,.
\end{align}

\subsubsection{\texorpdfstring{$\Gamma_3 \simeq A_4$}{Γ3=A4}}
\begin{align}
\mathbf{1}:&\quad \rho(S)= 1 ,\quad \rho(T)= 1\,, \\
\mathbf{1'}:&\quad \rho(S)= 1,\quad \rho(T)= \omega\,,\\
\mathbf{1''}:&\quad \rho(S)= 1,\quad \rho(T)= \omega^2\,,\\
\mathbf{3}:&\quad \rho(S)= 
\frac{1}{3}
\begin{pmatrix}
-1 & 2 & 2\\
2 & -1 & 2\\
2 & 2 & -1
\end{pmatrix}
,\quad \rho(T)= 
\begin{pmatrix}
1 & 0 & 0 \\
0 & \omega & 0 \\ 
0 & 0 & \omega^2
\end{pmatrix}\,,
\end{align}
where $\omega = e^{2\pi i/3}$.

\subsubsection{\texorpdfstring{$\Gamma_4 \simeq S_4$}{Γ4=S4}}
\begin{align}
\mathbf{1}:&\quad \rho(S)= 1 ,\quad \rho(T)= 1\,, \\
\mathbf{1'}:&\quad \rho(S)= -1,\quad \rho(T)= -1\,,\\
\mathbf{2}:&\quad \rho(S)= 
\frac{1}{2}
\begin{pmatrix}
-1 & \sqrt{3} \\ \sqrt{3} & 1
\end{pmatrix}
,\quad \rho(T)= 
\begin{pmatrix}
1 & 0 \\ 0 & -1 
\end{pmatrix}
\,,  \\
\mathbf{3}:&\quad \rho(S)=
-\frac{1}{2}
\begin{pmatrix}
0 & \sqrt{2} &\sqrt{2}\\
\sqrt{2} & -1 & 1 \\
\sqrt{2} & 1 & -1
\end{pmatrix} 
,\quad \rho(T)=
-
\begin{pmatrix}
1 & 0 & 0 \\
0 & i & 0 \\
0 & 0 & -i
\end{pmatrix}  \,,\\
\mathbf{3'}:&\quad \rho(S)=
\frac{1}{2}
\begin{pmatrix}
0 & \sqrt{2} &\sqrt{2}\\
\sqrt{2} & -1 & 1 \\
\sqrt{2} & 1 & -1
\end{pmatrix} 
,\quad \rho(T)=
\begin{pmatrix}
1 & 0 & 0 \\
0 & i & 0 \\
0 & 0 & -i
\end{pmatrix} \,.
\end{align}

\subsubsection{\texorpdfstring{$\Gamma_5 \simeq A_5$}{Γ5=A5}}
\begin{align}
\mathbf{1}:&\quad \rho(S)= 1 ,\quad \rho(T)= 1\,,\\
\mathbf{3}:&\quad \rho(S)=
\frac{1}{\sqrt{5}}
\begin{pmatrix}
 1 & -\sqrt{2} & -\sqrt{2} \\
 -\sqrt{2} & -\varphi & 1/{\varphi} \\
 -\sqrt{2} & 1/{\varphi} & -\varphi
\end{pmatrix} 
,\quad \rho(T)=
\begin{pmatrix}
1 & 0 & 0 \\
0 & \zeta & 0 \\
0 & 0 & \zeta^4   
\end{pmatrix}  \,,\\
\mathbf{3'}:&\quad \rho(S)= 
\frac{1}{\sqrt{5}}
\begin{pmatrix}
 -1 & \sqrt{2} & \sqrt{2} \\
 \sqrt{2} & -1/\varphi & \varphi \\
 \sqrt{2} & \varphi & -1/\varphi
\end{pmatrix} 
,\quad \rho(T)=
\begin{pmatrix}
1 & 0 & 0 \\
0 & \zeta^2 & 0 \\
0 & 0 & \zeta^3   \end{pmatrix}\,,\\
\mathbf{4}:&\quad \rho(S)= 
\frac{1}{\sqrt{5}}
\begin{pmatrix}
 1 & 1/\varphi & \varphi & -1 \\
 1/\varphi & -1 & 1 & \varphi \\
 \varphi & 1 & -1 & 1/\varphi \\
 -1 & \varphi & 1/\varphi & 1 
\end{pmatrix} 
,\quad \rho(T)=
\begin{pmatrix}
\zeta & 0 & 0 & 0\\
0 & \zeta^2 & 0 &0 \\
0 & 0 & \zeta^3 & 0 \\
0 & 0 & 0 & \zeta^4 
\end{pmatrix}\,,\\
\mathbf{5}:&\quad \rho(S)= 
\frac{1}{5}
\begin{pmatrix}
-1 & \sqrt{6} & \sqrt{6} & \sqrt{6} & \sqrt{6} \\
 \sqrt{6} & 1/{\varphi^2} & -2 \varphi  & {2}/{\varphi } & \varphi ^2 \\
 \sqrt{6} & -2 \varphi  & \varphi ^2 & {1}/{\varphi ^2} & {2}/{\varphi } \\
 \sqrt{6} & {2}/{\varphi } & 1/{\varphi ^2} & \varphi ^2 & -2 \varphi  \\
 \sqrt{6} & \varphi ^2 & {2}/{\varphi } & -2 \varphi  & {1}/{\varphi ^2}
\end{pmatrix} 
,\quad \rho(T)=
\begin{pmatrix}
1 & 0 & 0 & 0 & 0\\
0 & \zeta & 0 & 0 & 0\\
0 & 0 & \zeta^2 & 0 &0 \\
0 & 0 & 0 & \zeta^3 & 0 \\
0 & 0 & 0 & 0 & \zeta^4 
\end{pmatrix}\,, 
\end{align}
where $\zeta = e^{2\pi i/5}$ and $\varphi = (1+\sqrt{5})/2$.

%=======================
\subsection{Clebsch-Gordan Coefficients}
\label{app:CGCs}
%=======================
For completeness, and for each level $N = 2,3,4,5$, we also reproduce here the nontrivial Clebsch-Gordan coefficients
in the symmetric basis of Appendix~\ref{app:sym_basis}.
Entries of each multiplet entering the tensor product are denoted by $\alpha_i$ and $\beta_i$.
\subsubsection{\texorpdfstring{$\Gamma_2 \simeq S_3$}{Γ2=S3}}
%%%%%%%%%%%%%%%%%
\begin{align}
\begin{array}{@{}c@{{}\,\otimes\,{}}c@{{}\,\,=\,\,{}}l@{\quad\sim\quad}l@{}}
\mathbf{1'}&\mathbf{1'}&\mathbf{1}&\alpha_1\,\beta_1 \\[2mm]
\mathbf{1'}&\mathbf{2}&\mathbf{2}&
\begin{pmatrix}
-\alpha_1\,\beta_2\\
\alpha_1\,\beta_1 
\end{pmatrix}
\end{array}
\end{align}
%%%%%%%%%%%%%%%%%
%
%
%
%%%%%%%%%%%%%%%%%
\begin{align}
\begin{array}{@{}c@{{}\,\otimes\,{}}c@{{}\,\,=\,\,{}}ll@{}}
\mathbf{2}&\mathbf{2}&\mathbf{1}\,\oplus\, \mathbf{1'}\,\oplus\, \mathbf{2} &
\left\{\begin{array}{@{}l@{\quad\sim\quad}l@{}}
\quad \mathbf{1}  & \alpha_1\beta_1+\alpha_2\beta_2\\[2mm]
\quad \mathbf{1'} & \alpha_1\beta_2-\alpha_2\beta_1\\[2mm]
\quad \mathbf{2}  & \begin{pmatrix} \alpha_1\,\beta_2 + \alpha_2\,\beta_1 \\
                                    \alpha_1\,\beta_1 - \alpha_2\,\beta_2
                    \end{pmatrix}
\end{array}\right.
\end{array}
\end{align}
%%%%%%%%%%%%%%%%%
%

\subsubsection{\texorpdfstring{$\Gamma_3 \simeq A_4$}{Γ3=A4}}
%%%%%%%%%%%%%%%%%
\begin{align}
\begin{array}{@{}c@{{}\,\otimes\,{}}c@{{}\,\,=\,\,{}}l@{\quad\sim\quad}l@{}}
\mathbf{1'}&\mathbf{1'}&\mathbf{1''}&\alpha_1\,\beta_1 \\[2mm]
\mathbf{1''}&\mathbf{1''}&\mathbf{1'}&\alpha_1\,\beta_1 \\[2mm]
\mathbf{1'}&\mathbf{1''}&\mathbf{1}&\alpha_1\,\beta_1 \\[4mm]
\mathbf{1'}&\mathbf{3}&\mathbf{3}&
\begin{pmatrix}
\alpha_1\,\beta_3 \\
\alpha_1\,\beta_1 \\
\alpha_1\,\beta_2
\end{pmatrix} \\[6mm]
\mathbf{1''}&\mathbf{3}&\mathbf{3}&
\begin{pmatrix}
\alpha_1\,\beta_2 \\
\alpha_1\,\beta_3 \\
\alpha_1\,\beta_1
\end{pmatrix} 
\end{array}
\end{align}
%%%%%%%%%%%%%%%%%
%
%
%
%%%%%%%%%%%%%%%%%
\begin{align}
\begin{array}{@{}c@{{}\,\otimes\,{}}c@{{}\,\,=\,\,{}}ll@{}}
\mathbf{3}&\mathbf{3}&\mathbf{1}\,\oplus\,\mathbf{1'}\,\oplus\,\mathbf{1''}\,\oplus\,\mathbf{3}_1\,\oplus\,\mathbf{3}_2&
\left\{\begin{array}{@{}l@{\quad\sim\quad}l@{}}
\quad \mathbf{1}    & \alpha_1\beta_1+\alpha_2\beta_3+\alpha_3\beta_2\\[2mm]
\quad \mathbf{1'}   & \alpha_1\beta_2+\alpha_2\beta_1+\alpha_3\beta_3\\[2mm]
\quad \mathbf{1''}  & \alpha_1\beta_3+\alpha_2\beta_2+\alpha_3\beta_1\\[2mm]
\quad \mathbf{3}_1  & \begin{pmatrix} 2 \alpha_1\,\beta_1-\alpha_2\,\beta_3-\alpha_3\,\beta_2\\
                                      2 \alpha_3\,\beta_3-\alpha_1\,\beta_2-\alpha_2\,\beta_1\\
                                      2 \alpha_2\,\beta_2-\alpha_3\,\beta_1-\alpha_1\,\beta_3
                    \end{pmatrix}\\[6mm]
\quad \mathbf{3}_2  & \begin{pmatrix} \alpha_2\,\beta_3-\alpha_3\,\beta_2\\
                                    \alpha_1\,\beta_2-\alpha_2\,\beta_1\\
                                    \alpha_3\,\beta_1-\alpha_1\,\beta_3
                    \end{pmatrix}\\[6mm]
\end{array}\right.
\end{array}
\end{align}
%%%%%%%%%%%%%%%%%
%

\subsubsection{\texorpdfstring{$\Gamma_4 \simeq S_4$}{Γ4=S4}}
%%%%%%%%%%%%%%%%%
\begin{align}
\begin{array}{@{}c@{{}\,\otimes\,{}}c@{{}\,\,=\,\,{}}l@{\quad\sim\quad}l@{}}
\mathbf{1'}&\mathbf{1'}&\mathbf{1}&\alpha_1\,\beta_1 \\[2mm]
\mathbf{1'}&\mathbf{2}&\mathbf{2}&
\begin{pmatrix}
\alpha_1\,\beta_2\\
-\alpha_1\,\beta_1 
\end{pmatrix}\\[4mm]
\mathbf{1'}&\mathbf{3}&\mathbf{3'}&
\begin{pmatrix}
\alpha_1\,\beta_1 \\
\alpha_1\,\beta_2 \\
\alpha_1\,\beta_3
\end{pmatrix} \\[6mm]
\mathbf{1'}&\mathbf{3'}&\mathbf{3}&
\begin{pmatrix}
\alpha_1\,\beta_1 \\
\alpha_1\,\beta_2 \\
\alpha_1\,\beta_3
\end{pmatrix} 
\end{array}
\end{align}
%%%%%%%%%%%%%%%%%
%
\vskip 5mm
%
%%%%%%%%%%%%%%%%%
\begin{align}
\begin{array}{@{}c@{{}\,\otimes\,{}}c@{{}\,\,=\,\,{}}ll@{}}
\mathbf{2}&\mathbf{2}&\mathbf{1}\,\oplus\, \mathbf{1'}\,\oplus\, \mathbf{2} &
\left\{\begin{array}{@{}l@{\quad\sim\quad}l@{}}
\quad \mathbf{1}  & \alpha_1\beta_1+\alpha_2\beta_2\\[2mm]
\quad \mathbf{1'} & \alpha_1\beta_2-\alpha_2\beta_1\\[2mm]
\quad \mathbf{2}  & \begin{pmatrix} \alpha_2\,\beta_2 - \alpha_1\,\beta_1\\
                                    \alpha_1\,\beta_2 + \alpha_2\,\beta_1
                    \end{pmatrix}
\end{array}\right.
\\[13mm]
\mathbf{2}&\mathbf{3}&\mathbf{3}\,\oplus\,\mathbf{3'}&
\left\{\begin{array}{@{}l@{\quad\sim\quad}l@{}}
\quad \mathbf{3}   & \begin{pmatrix} \alpha_1\,\beta_1 \\
                                     \left(\sqrt{3}/2\right)\alpha_2\,\beta_3 - \left(1/2\right)\alpha_1\,\beta_2 \\
                                     \left(\sqrt{3}/2\right)\alpha_2\,\beta_2 - \left(1/2\right)\alpha_1\,\beta_3
                     \end{pmatrix}\\[6mm]
\quad \mathbf{3'}  & \begin{pmatrix} - \alpha_2\,\beta_1 \\
                                     \left(\sqrt{3}/2\right)\alpha_1\,\beta_3 + \left(1/2\right)\alpha_2\,\beta_2 \\
                                     \left(\sqrt{3}/2\right)\alpha_1\,\beta_2 + \left(1/2\right)\alpha_2\,\beta_3
                     \end{pmatrix}
\end{array}\right.
\\[16mm]
\mathbf{2}&\mathbf{3'}&\mathbf{3}\,\oplus\,\mathbf{3'}&
\left\{\begin{array}{@{}l@{\quad\sim\quad}l@{}}
\quad \mathbf{3}   & \begin{pmatrix} - \alpha_2\,\beta_1 \\
                                     \left(\sqrt{3}/2\right)\alpha_1\,\beta_3 + \left(1/2\right)\alpha_2\,\beta_2 \\
                                     \left(\sqrt{3}/2\right)\alpha_1\,\beta_2 + \left(1/2\right)\alpha_2\,\beta_3
                     \end{pmatrix}\\[6mm]
\quad \mathbf{3'}  & \begin{pmatrix} \alpha_1\,\beta_1 \\
                                     \left(\sqrt{3}/2\right)\alpha_2\,\beta_3 - \left(1/2\right)\alpha_1\,\beta_2 \\
                                     \left(\sqrt{3}/2\right)\alpha_2\,\beta_2 - \left(1/2\right)\alpha_1\,\beta_3
                     \end{pmatrix}
\end{array}\right.
\end{array}
\end{align}
%%%%%%%%%%%%%%%%%
%
\vfill\pagebreak
%
%%%%%%%%%%%%%%%%%
\begin{align}
\begin{array}{@{}cll@{}}
\mathbf{3}\,\otimes\,\mathbf{3}\,=\,\mathbf{3'}\,\otimes\,\mathbf{3'}\,=\,\mathbf{1}\,\oplus\, \mathbf{2}\,\oplus\, \mathbf{3}\,\oplus\, \mathbf{3'} &
\left\{\begin{array}{@{}l@{\quad\sim\quad}l@{}}
\quad \mathbf{1}  & \alpha_1\beta_1+\alpha_2\beta_3+\alpha_3\beta_2\\[2mm]
\quad \mathbf{2}  & \begin{pmatrix}
                      \alpha_1\beta_1- \left(1/2\right)\left(\alpha_2\beta_3+\alpha_3\beta_2\right) \\
                    \left(\sqrt{3}/2\right) \left(\alpha_2\beta_2+\alpha_3\beta_3\right) 
                    \end{pmatrix}\\[4mm]
\quad \mathbf{3}   & \begin{pmatrix} 
                       \alpha_3\beta_3-\alpha_2\beta_2 \\
                       \alpha_1\beta_3+\alpha_3\beta_1 \\
                       -\alpha_1\beta_2-\alpha_2\beta_1
                     \end{pmatrix}\\[6mm]
\quad \mathbf{3'}  & \begin{pmatrix}
                       \alpha_3\beta_2-\alpha_2\beta_3 \\
                       \alpha_2\beta_1-\alpha_1\beta_2\\
                       \alpha_1\beta_3-\alpha_3\beta_1
                     \end{pmatrix}
\end{array}\right.
\end{array}
\end{align}
%%%%%%%%%%%%%%%%%
%
%
%
%%%%%%%%%%%%%%%%%
\begin{align}
\begin{array}{@{}cll@{}}
\mathbf{3}\,\otimes\,\mathbf{3'}\,=\,\mathbf{1'}\,\oplus\, \mathbf{2}\,\oplus\, \mathbf{3}\,\oplus\, \mathbf{3'} &
\left\{\begin{array}{@{}l@{\quad\sim\quad}l@{}}
\quad \mathbf{1'} & \alpha_1\beta_1+\alpha_2\beta_3+\alpha_3\beta_2\\[2mm]
\quad \mathbf{2}  & \begin{pmatrix}
                      \left(\sqrt{3}/2\right) \left(\alpha_2\beta_2+\alpha_3\beta_3\right)  \\
                      -\alpha_1\beta_1+ \left(1/2\right)\left(\alpha_2\beta_3+\alpha_3\beta_2\right)                   
                    \end{pmatrix}\\[4mm]
\quad \mathbf{3}   & \begin{pmatrix} 
                       \alpha_3\beta_2-\alpha_2\beta_3 \\
                       \alpha_2\beta_1-\alpha_1\beta_2\\
                       \alpha_1\beta_3-\alpha_3\beta_1
                     \end{pmatrix}\\[6mm]
\quad \mathbf{3'}  & \begin{pmatrix}
                       \alpha_3\beta_3-\alpha_2\beta_2 \\
                       \alpha_1\beta_3+\alpha_3\beta_1 \\
                       -\alpha_1\beta_2-\alpha_2\beta_1
                     \end{pmatrix}
\end{array}\right.
\end{array}
\end{align}
%%%%%%%%%%%%%%%%%
%

\vskip 2cm

\subsubsection{\texorpdfstring{$\Gamma_5 \simeq A_5$}{Γ5=A5}}
%%%%%%%%%%%%%%
\begin{align}
\begin{array}{@{}c@{{}\,\otimes\,{}}c@{{}\,\,=\,\,{}}ll@{}}
\mathbf{3}&\mathbf{3}&\mathbf{1}\,\oplus\, \mathbf{3}\,\oplus\, \mathbf{5} &
\left\{\begin{array}{@{}l@{\quad\sim\quad}l@{}}
\quad \mathbf{1}  & \alpha_1\beta_1+\alpha_2\beta_3+\alpha_3\beta_2\\[2mm]
\quad \mathbf{3}  & \begin{pmatrix} \alpha_2\,\beta_3-\alpha_3\,\beta_2\\
                                    \alpha_1\,\beta_2-\alpha_2\,\beta_1\\
                                    \alpha_3\,\beta_1-\alpha_1\,\beta_3
                    \end{pmatrix}\\[6mm]
\quad \mathbf{5}  & \begin{pmatrix}  2 \alpha_1 \beta_1-\alpha_2 \beta_3-\alpha_3 \beta_2 \\
                                     -\sqrt{3}\, \alpha_1 \beta_2-\sqrt{3}\, \alpha_2 \beta_1 \\
                                     \sqrt{6}\, \alpha_2 \beta_2 \\
                                     \sqrt{6}\, \alpha_3 \beta_3 \\
                                     -\sqrt{3}\, \alpha_1 \beta_3-\sqrt{3}\, \alpha_3 \beta_1
                    \end{pmatrix}
\end{array}\right.
\end{array}
\end{align}
%%%%%%%%%%%%%%
%
%
%%%%%%%%%%%%%%
\begin{align}
\begin{array}{@{}c@{{}\,\otimes\,{}}c@{{}\,\,=\,\,{}}ll@{}}
\mathbf{3}&\mathbf{3'}&\mathbf{4}\,\oplus\,\mathbf{5}&
\left\{\begin{array}{@{}l@{\quad\sim\quad}l@{}}
\quad \mathbf{4}   & \begin{pmatrix}  \sqrt{2}\, \alpha_2 \beta_1+\alpha_3 \beta_2 \\
                                     -\sqrt{2}\, \alpha_1 \beta_2-\alpha_3 \beta_3 \\
                                     -\sqrt{2}\, \alpha_1 \beta_3-\alpha_2 \beta_2 \\
                                      \sqrt{2}\, \alpha_3 \beta_1+\alpha_2 \beta_3
                     \end{pmatrix}\\[9mm]
\quad \mathbf{5}  & \begin{pmatrix} \sqrt{3}\, \alpha_1 \beta_1 \\
                                    \alpha_2 \beta_1-\sqrt{2}\, \alpha_3 \beta_2 \\
                                    \alpha_1 \beta_2-\sqrt{2}\, \alpha_3 \beta_3 \\
                                    \alpha_1 \beta_3-\sqrt{2}\, \alpha_2 \beta_2 \\
                                    \alpha_3 \beta_1-\sqrt{2}\, \alpha_2 \beta_3
                     \end{pmatrix}
\end{array}\right.
\\[23mm]
\mathbf{3}&\mathbf{4}&\mathbf{3'}\,\oplus\,\mathbf{4}\,\oplus\,\mathbf{5}&
\left\{\begin{array}{@{}l@{\quad\sim\quad}l@{}}
\quad \mathbf{3'}   & \begin{pmatrix}  -\sqrt{2}\, \alpha_2 \beta_4-\sqrt{2}\, \alpha_3 \beta_1 \\
                                        \sqrt{2}\, \alpha_1 \beta_2-\alpha_2 \beta_1+\alpha_3 \beta_3 \\
                                        \sqrt{2}\,\alpha_1 \beta_3+\alpha_2 \beta_2-\alpha_3 \beta_4
                     \end{pmatrix}\\[6mm]
\quad \mathbf{4}   & \begin{pmatrix}   \alpha_1 \beta_1-\sqrt{2}\,\alpha_3 \beta_2 \\
                                      -\alpha_1 \beta_2-\sqrt{2}\,\alpha_2 \beta_1 \\
                                       \alpha_1 \beta_3+\sqrt{2}\,\alpha_3 \beta_4 \\
                                      -\alpha_1 \beta_4+\sqrt{2}\,\alpha_2 \beta_3
                     \end{pmatrix}\\[9mm]
\quad \mathbf{5}   & \begin{pmatrix}  \sqrt{6}\,\alpha_2\beta_4-\sqrt{6}\, \alpha_3 \beta_1 \\
                                       2\sqrt{2}\,\alpha_1 \beta_1+2 \alpha_3 \beta_2\\
                                       -\sqrt{2}\,\alpha_1 \beta_2+\alpha_2 \beta_1+3\alpha_3 \beta_3 \\
                                      \sqrt{2}\, \alpha_1 \beta_3-3\alpha_2 \beta_2-\alpha_3 \beta_4\\
                                       -2\sqrt{2}\, \alpha_1 \beta_4-2 \alpha_2 \beta_3
                     \end{pmatrix}
\end{array}\right.
\\[30mm]
\mathbf{3}&\mathbf{5}&\mathbf{3}\,\oplus\,\mathbf{3'}\,\oplus\,\mathbf{4}\,\oplus\,\mathbf{5}&
\left\{\begin{array}{@{}l@{\quad\sim\quad}l@{}}
\quad \mathbf{3}   & \begin{pmatrix}
                          -2 \alpha_1 \beta_1+\sqrt{3}\,\alpha_2 \beta_5+\sqrt{3}\,\alpha_3 \beta_2 \\
                          \sqrt{3}\,\alpha_1 \beta_2+\alpha_2 \beta_1-\sqrt{6}\,\alpha_3 \beta_3 \\
                          \sqrt{3}\,\alpha_1 \beta_5-\sqrt{6}\,\alpha_2 \beta_4+\alpha_3 \beta_1
                     \end{pmatrix}\\[6mm]
\quad \mathbf{3'}   & \begin{pmatrix} 
                          \sqrt{3}\,\alpha_1 \beta_1+\alpha_2 \beta_5+\alpha_3 \beta_2 \\
                          \alpha_1 \beta_3-\sqrt{2}\,\alpha_2 \beta_2-\sqrt{2}\,\alpha_3 \beta_4 \\
                          \alpha_1 \beta_4-\sqrt{2}\,\alpha_2 \beta_3-\sqrt{2}\,\alpha_3 \beta_5
                     \end{pmatrix}\\[6mm]
\quad \mathbf{4}   & \begin{pmatrix}
                          2\sqrt{2}\,\alpha_1 \beta_2-\sqrt{6}\, \alpha_2 \beta_1+\alpha_3 \beta_3 \\
                          -\sqrt{2}\,\alpha_1 \beta_3+2\alpha_2 \beta_2-3 \alpha_3 \beta_4 \\
                           \sqrt{2}\,\alpha_1 \beta_4+3\alpha_2 \beta_3-2\alpha_3 \beta_5 \\
                          -2\sqrt{2}\,\alpha_1 \beta_5-\alpha_2 \beta_4+\sqrt{6}\,\alpha_3 \beta_1
                     \end{pmatrix}\\[9mm]
\quad \mathbf{5}   & \begin{pmatrix}
                          \sqrt{3}\, \alpha_2 \beta_5-\sqrt{3}\, \alpha_3 \beta_2 \\
                          -\alpha_1 \beta_2-\sqrt{3}\,\alpha_2 \beta_1-\sqrt{2}\,\alpha_3 \beta_3 \\
                          -2 \alpha_1 \beta_3-\sqrt{2}\,\alpha_2 \beta_2 \\
                          2\alpha_1 \beta_4+\sqrt{2}\,\alpha_3 \beta_5 \\
                          \alpha_1 \beta_5+\sqrt{2}\,\alpha_2 \beta_4+ \sqrt{3}\,\alpha_3 \beta_1
                     \end{pmatrix}
\end{array}\right.
\end{array}
\end{align}

\begin{align}
\begin{array}{@{}c@{{}\,\otimes\,{}}c@{{}\,\,=\,\,{}}ll@{}}
\mathbf{3'}&\mathbf{3'}&\mathbf{1}\,\oplus\,\mathbf{3'}\,\oplus\,\mathbf{5}&
\left\{\begin{array}{@{}l@{\quad\sim\quad}l@{}}
\quad \mathbf{1}  & \alpha_1\beta_1+\alpha_2\beta_3+\alpha_3\beta_2\\[2mm]
\quad \mathbf{3'}  & \begin{pmatrix} \alpha_2\,\beta_3-\alpha_3\,\beta_2\\
                                    \alpha_1\,\beta_2-\alpha_2\,\beta_1\\
                                    \alpha_3\,\beta_1-\alpha_1\,\beta_3
                    \end{pmatrix}\\[6mm]
\quad \mathbf{5}  & \begin{pmatrix}  2 \alpha_1 \beta_1-\alpha_2 \beta_3-\alpha_3 \beta_2 \\
                                     \sqrt{6}\, \alpha_3 \beta_3 \\
                                     -\sqrt{3}\, \alpha_1 \beta_2-\sqrt{3}\, \alpha_2 \beta_1 \\
                                     -\sqrt{3}\, \alpha_1 \beta_3-\sqrt{3}\, \alpha_3 \beta_1 \\
                                     \sqrt{6}\, \alpha_2 \beta_2
                    \end{pmatrix}
\end{array}\right.
\\[23mm]
\mathbf{3'}&\mathbf{4}&\mathbf{3}\,\oplus\,\mathbf{4}\,\oplus\,\mathbf{5}&
\left\{\begin{array}{@{}l@{\quad\sim\quad}l@{}}
\quad \mathbf{3}   & \begin{pmatrix} -\sqrt{2}\,\alpha_2 \beta_3-\sqrt{2}\,\alpha_3 \beta_2 \\
                                     \sqrt{2}\,\alpha_1 \beta_1+\alpha_2 \beta_4-\alpha_3 \beta_3 \\
                                     \sqrt{2}\,\alpha_1 \beta_4 -\alpha_2 \beta_2+\alpha_3 \beta_1
                     \end{pmatrix}\\[6mm]
\quad \mathbf{4}   & \begin{pmatrix} \alpha_1 \beta_1+\sqrt{2}\,\alpha_3 \beta_3 \\
                                     \alpha_1 \beta_2-\sqrt{2}\,\alpha_3 \beta_4 \\
                                    -\alpha_1 \beta_3+\sqrt{2}\,\alpha_2 \beta_1 \\
                                    -\alpha_1 \beta_4-\sqrt{2}\,\alpha_2 \beta_2
                     \end{pmatrix}\\[9mm]
\quad \mathbf{5}   & \begin{pmatrix}  \sqrt{6}\,\alpha_2 \beta_3-\sqrt{6}\,\alpha_3 \beta_2 \\
                                      \sqrt{2}\,\alpha_1 \beta_1-3\alpha_2 \beta_4-\alpha_3 \beta_3 \\
                                      2\sqrt{2}\,\alpha_1 \beta_2+2 \alpha_3 \beta_4 \\
                                     -2\sqrt{2}\,\alpha_1 \beta_3-2\alpha_2 \beta_1 \\
                                     -\sqrt{2}\,\alpha_1 \beta_4+\alpha_2 \beta_2+3\alpha_3 \beta_1
                     \end{pmatrix}
\end{array}\right.
\\[30mm]
\mathbf{3'}&\mathbf{5}&\mathbf{3}\,\oplus\,\mathbf{3'}\,\oplus\,\mathbf{4}\,\oplus\,\mathbf{5}&
\left\{\begin{array}{@{}l@{\quad\sim\quad}l@{}}
\quad \mathbf{3}   & \begin{pmatrix}
              \sqrt{3}\, \alpha_1 \beta_1+\alpha_2\beta_4+\alpha_3 \beta_3 \\
              \alpha_1 \beta_2-\sqrt{2}\,\alpha_2 \beta_5 -\sqrt{2}\,\alpha_3 \beta_4\\
              \alpha_1 \beta_5-\sqrt{2}\,\alpha_2 \beta_3-\sqrt{2}\,\alpha_3 \beta_2
                     \end{pmatrix}\\[6mm]
\quad \mathbf{3'}   & \begin{pmatrix} 
              -2 \alpha_1 \beta_1+\sqrt{3}\,\alpha_2 \beta_4 +\sqrt{3}\,\alpha_3 \beta_3\\
              \sqrt{3}\,\alpha_1 \beta_3+\alpha_2 \beta_1-\sqrt{6}\,\alpha_3 \beta_5 \\
              \sqrt{3}\,\alpha_1 \beta_4-\sqrt{6}\,\alpha_2 \beta_2+\alpha_3 \beta_1
                     \end{pmatrix}\\[6mm]
\quad \mathbf{4}   & \begin{pmatrix}
              \sqrt{2}\,\alpha_1 \beta_2+3 \alpha_2\beta_5-2\alpha_3 \beta_4 \\
              2\sqrt{2}\,\alpha_1 \beta_3-\sqrt{6}\,\alpha_2 \beta_1+\alpha_3 \beta_5 \\
              -2\sqrt{2}\,\alpha_1 \beta_4-\alpha_2 \beta_2 +\sqrt{6}\,\alpha_3 \beta_1\\
              -\sqrt{2}\,\alpha_1\beta_5+2 \alpha_2\beta_3-3\alpha_3 \beta_2
                     \end{pmatrix}\\[9mm]
\quad \mathbf{5}   & \begin{pmatrix}
              \sqrt{3}\,\alpha_2 \beta_4-\sqrt{3} \alpha_3 \beta_3 \\
              2 \alpha_1 \beta_2+\sqrt{2}\,\alpha_3 \beta_4 \\
              -\alpha_1 \beta_3-\sqrt{3}\,\alpha_2 \beta_1-\sqrt{2}\,\alpha_3 \beta_5 \\
              \alpha_1 \beta_4+\sqrt{2}\,\alpha_2 \beta_2 + \sqrt{3}\,\alpha_3 \beta_1\\
              -2\alpha_1 \beta_5-\sqrt{2}\,\alpha_2 \beta_3
                     \end{pmatrix}
\end{array}\right.
\end{array}
\end{align}
%%%%%%%%%%%%%%
%
%
%
%%%%%%%%%%%%%%
\begin{align}
\begin{array}{@{}cll@{}}
\makecell{\mathbf{4}\,\otimes\,\mathbf{4}{{}\,\,=\,\,{}}\\
\mathbf{1}\,\oplus\,\mathbf{3}\,\oplus\,\mathbf{3'}\\\,\oplus\,\mathbf{4}\,\oplus\,\mathbf{5}}&
\left\{\begin{array}{@{}l@{\quad\sim\quad}l@{}}
\quad \mathbf{1}  & \alpha_1\beta_4+\alpha_2 \beta_3+\alpha_3 \beta_2+\alpha_4 \beta_1 \\[2mm]
\quad \mathbf{3}  & \begin{pmatrix}
              -\alpha_1 \beta_4+\alpha_2\beta_3-\alpha_3\beta_2+\alpha_4 \beta_1\\
              \sqrt{2}\,\alpha_2 \beta_4-\sqrt{2}\,\alpha_4 \beta_2\\
              \sqrt{2}\,\alpha_1 \beta_3-\sqrt{2}\,\alpha_3 \beta_1
                    \end{pmatrix}\\[6mm]
\quad \mathbf{3'}   & \begin{pmatrix}
              \alpha_1 \beta_4 +\alpha_2 \beta_3-\alpha_3 \beta_2 -\alpha_4 \beta_1\\
              \sqrt{2}\,\alpha_3 \beta_4-\sqrt{2}\,\alpha_4 \beta_3 \\
              \sqrt{2}\,\alpha_1 \beta_2-\sqrt{2}\,\alpha_2 \beta_1
                     \end{pmatrix}\\[6mm]
\quad \mathbf{4}   & \begin{pmatrix} 
              \alpha_2 \beta_4+\alpha_3\beta_3+\alpha_4 \beta_2 \\
              \alpha_1 \beta_1+\alpha_3 \beta_4 +\alpha_4 \beta_3\\
              \alpha_1\beta_2+\alpha_2 \beta_1+\alpha_4 \beta_4 \\
              \alpha_1 \beta_3+\alpha_2\beta_2+\alpha_3 \beta_1
                     \end{pmatrix}\\[9mm]
\quad \mathbf{5}   & \begin{pmatrix}
              \sqrt{3}\,\alpha_1 \beta_4-\sqrt{3}\,\alpha_2 \beta_3-\sqrt{3}\,\alpha_3 \beta_2+\sqrt{3}\,\alpha_4 \beta_1 \\
              -\sqrt{2}\,\alpha_2 \beta_4+2\sqrt{2}\,\alpha_3 \beta_3-\sqrt{2}\,\alpha_4 \beta_2 \\
              -2 \sqrt{2}\,\alpha_1 \beta_1+\sqrt{2}\,\alpha_3 \beta_4+\sqrt{2}\,\alpha_4 \beta_3 \\
              \sqrt{2}\,\alpha_1 \beta_2+\sqrt{2}\,\alpha_2 \beta_1-2 \sqrt{2}\,\alpha_4 \beta_4 \\
              -\sqrt{2}\,\alpha_1 \beta_3+2\sqrt{2}\,\alpha_2 \beta_2-\sqrt{2}\,\alpha_3 \beta_1
                     \end{pmatrix}
\end{array}\right.
\\[41mm]
\makecell{\mathbf{4}\,\otimes\,\mathbf{5}{{}\,\,=\,\,{}}\\
\mathbf{3}\,\oplus\,\mathbf{3'}\,\oplus\,\mathbf{4}\\\,\oplus\,\mathbf{5}_1\,\oplus\,\mathbf{5}_2}&
\left\{\begin{array}{@{}l@{\quad\sim\quad}l@{}}
\quad \mathbf{3}   & \begin{pmatrix} 
              2 \sqrt{2}\,\alpha_1\beta_5-\sqrt{2}\,\alpha_2 \beta_4+\sqrt{2}\,\alpha_3 \beta_3-2 \sqrt{2}\,\alpha_4 \beta_2\\
              -\sqrt{6}\,\alpha_1 \beta_1+2 \alpha_2 \beta_5+3 \alpha_3 \beta_4-\alpha_4 \beta_3 \\
              \alpha_1 \beta_4-3 \alpha_2\beta_3-2\alpha_3 \beta_2+\sqrt{6}\,\alpha_4 \beta_1
                     \end{pmatrix}\\[6mm]
\quad \mathbf{3'}   & \begin{pmatrix} 
              \sqrt{2}\,\alpha_1 \beta_5+2\sqrt{2}\,\alpha_2 \beta_4-2\sqrt{2}\,\alpha_3 \beta_3-\sqrt{2}\,\alpha_4 \beta_2 \\
              3\alpha_1 \beta_2-\sqrt{6}\, \alpha_2 \beta_1-\alpha_3 \beta_5+2 \alpha_4\beta_4 \\
              -2 \alpha_1 \beta_3+\alpha_2 \beta_2+\sqrt{6}\,\alpha_3 \beta_1-3 \alpha_4 \beta_5
                     \end{pmatrix}\\[6mm]
\quad \mathbf{4}   & \begin{pmatrix}
              \sqrt{3}\,\alpha_1 \beta_1-\sqrt{2}\,\alpha_2 \beta_5+\sqrt{2}\,\alpha_3 \beta_4-2\sqrt{2}\,\alpha_4 \beta_3 \\
              -\sqrt{2}\,\alpha_1 \beta_2-\sqrt{3}\,\alpha_2 \beta_1+2 \sqrt{2}\,\alpha_3 \beta_5+\sqrt{2}\,\alpha_4 \beta_4 \\
              \sqrt{2}\,\alpha_1 \beta_3+2\sqrt{2}\,\alpha_2 \beta_2-\sqrt{3}\,\alpha_3 \beta_1-\sqrt{2}\,\alpha_4 \beta_5\\
              -2 \sqrt{2}\,\alpha_1 \beta_4+\sqrt{2}\,\alpha_2 \beta_3-\sqrt{2}\,\alpha_3 \beta_2+\sqrt{3}\,\alpha_4 \beta_1 
                     \end{pmatrix}\\[9mm]
\quad \mathbf{5}_1   & \begin{pmatrix}
              \sqrt{2}\,\alpha_1 \beta_5-\sqrt{2}\,\alpha_2 \beta_4-\sqrt{2}\,\alpha_3 \beta_3+\sqrt{2}\,\alpha_4 \beta_2\\
              -\sqrt{2}\,\alpha_1 \beta_1-\sqrt{3}\,\alpha_3 \beta_4 -\sqrt{3}\,\alpha_4 \beta_3 \\
              \sqrt{3}\,\alpha_1 \beta_2+\sqrt{2}\,\alpha_2 \beta_1+\sqrt{3}\,\alpha_3 \beta_5 \\
              \sqrt{3}\,\alpha_2 \beta_2+\sqrt{2}\,\alpha_3 \beta_1+\sqrt{3}\,\alpha_4 \beta_5 \\
              -\sqrt{3}\,\alpha_1 \beta_4-\sqrt{3}\,\alpha_2 \beta_3-\sqrt{2}\,\alpha_4 \beta_1
                     \end{pmatrix}\\[11mm]
\quad \mathbf{5}_2   & \begin{pmatrix} 
              2 \alpha_1\beta_5+4 \alpha_2 \beta_4+4 \alpha_3 \beta_3 +2 \alpha_4 \beta_2\\
              4 \alpha_1 \beta_1+2 \sqrt{6}\,\alpha_2 \beta_5 \\
              -\sqrt{6}\,\alpha_1\beta_2+2 \alpha_2 \beta_1-\sqrt{6}\,\alpha_3 \beta_5 +2 \sqrt{6}\,\alpha_4 \beta_4\\
               2 \sqrt{6}\,\alpha_1 \beta_3-\sqrt{6}\,\alpha_2\beta_2+2 \alpha_3 \beta_1-\sqrt{6}\,\alpha_4 \beta_5 \\
               2 \sqrt{6}\,\alpha_3 \beta_2+4 \alpha_4 \beta_1
                     \end{pmatrix}
\end{array}\right.
\end{array}
\end{align}
%%%%%%%%%%%%%%
%
%
%
%%%%%%%%%%%%%%
\begin{align}
\begin{array}{@{}cll@{}}
\makecell{\mathbf{5}\,\otimes\,\mathbf{5}
{{}\,\,=\,\,{}}
\\ \mathbf{1}\,\oplus\,\mathbf{3}\,\oplus\,\mathbf{3'}\,\oplus\,\mathbf{4}_1\\\,\oplus\,\mathbf{4}_2\,\oplus\,\mathbf{5}_1\,\oplus\,\mathbf{5}_2}&
\left\{\begin{array}{@{}l@{\quad\sim\quad}l@{}}
\quad \mathbf{1}  & \alpha_1\beta_1+\alpha_2\beta_5+\alpha_3\beta_4+\alpha_4\beta_3+\alpha_5\beta_2 \\[2mm]
\quad \mathbf{3}  & \begin{pmatrix} 
     \alpha_2 \beta_5+2\alpha_3 \beta_4-2 \alpha_4 \beta_3-\alpha_5 \beta_2 \\
     -\sqrt{3}\,\alpha_1\beta_2+\sqrt{3}\,\alpha_2 \beta_1+\sqrt{2}\,\alpha_3 \beta_5-\sqrt{2}\,\alpha_5 \beta_3 \\
     \sqrt{3}\,\alpha_1 \beta_5+\sqrt{2}\,\alpha_2 \beta_4-\sqrt{2}\,\alpha_4\beta_2-\sqrt{3}\,\alpha_5 \beta_1
                    \end{pmatrix}\\[6mm]
\quad \mathbf{3'}   & \begin{pmatrix}
     2 \alpha_2 \beta_5-\alpha_3 \beta_4+\alpha_4 \beta_3-2 \alpha_5 \beta_2\\
     \sqrt{3}\,\alpha_1\beta_3-\sqrt{3}\,\alpha_3 \beta_1+\sqrt{2}\,\alpha_4 \beta_5-\sqrt{2}\,\alpha_5 \beta_4 \\
     -\sqrt{3}\,\alpha_1 \beta_4+\sqrt{2}\,\alpha_2 \beta_3-\sqrt{2}\,\alpha_3\beta_2+\sqrt{3}\,\alpha_4 \beta_1
                     \end{pmatrix}\\[6mm]
\quad \mathbf{4}_1   & \begin{pmatrix} 
     3 \sqrt{2}\,\alpha_1 \beta_2+3 \sqrt{2}\,\alpha_2 \beta_1-\sqrt{3}\,\alpha_3\beta_5+4 \sqrt{3}\,\alpha_4\beta_4-\sqrt{3}\,\alpha_5 \beta_3\\
     3 \sqrt{2}\,\alpha_1 \beta_3+4 \sqrt{3}\,\alpha_2 \beta_2+3 \sqrt{2}\,\alpha_3 \beta_1-\sqrt{3}\,\alpha_4\beta_5-\sqrt{3}\, \alpha_5 \beta_4 \\
     3 \sqrt{2}\,\alpha_1 \beta_4-\sqrt{3}\,\alpha_2 \beta_3-\sqrt{3}\,\alpha_3 \beta_2+3 \sqrt{2}\,\alpha_4 \beta_1+4\sqrt{3}\,\alpha_5\beta_5 \\
     3 \sqrt{2}\,\alpha_1\beta_5-\sqrt{3}\,\alpha_2 \beta_4+4 \sqrt{3}\,\alpha_3 \beta_3-\sqrt{3}\,\alpha_4 \beta_2+3\sqrt{2}\,\alpha_5 \beta_1
                     \end{pmatrix}\\[9mm]
\quad \mathbf{4}_2   & \begin{pmatrix}
     \sqrt{2}\,\alpha_1 \beta_2-\sqrt{2}\,\alpha_2 \beta_1+\sqrt{3}\,\alpha_3 \beta_5-\sqrt{3}\,\alpha_5 \beta_3 \\
     -\sqrt{2}\,\alpha_1 \beta_3+\sqrt{2}\,\alpha_3 \beta_1+\sqrt{3}\,\alpha_4 \beta_5-\sqrt{3}\,\alpha_5 \beta_4 \\
     -\sqrt{2}\,\alpha_1 \beta_4-\sqrt{3}\,\alpha_2 \beta_3+\sqrt{3}\,\alpha_3 \beta_2+\sqrt{2}\,\alpha_4 \beta_1\\
     \sqrt{2}\,\alpha_1 \beta_5-\sqrt{3}\,\alpha_2 \beta_4+\sqrt{3}\,\alpha_4 \beta_2-\sqrt{2}\,\alpha_5 \beta_1
                     \end{pmatrix}\\[9mm]
\quad \mathbf{5}_1   & \begin{pmatrix}
     2 \alpha_1 \beta_1+\alpha_2 \beta_5-2 \alpha_3 \beta_4-2 \alpha_4 \beta_3+\alpha_5 \beta_2 \\
     \alpha_1 \beta_2+\alpha_2 \beta_1+\sqrt{6}\,\alpha_3 \beta_5+\sqrt{6}\,\alpha_5 \beta_3 \\
     -2 \alpha_1 \beta_3+\sqrt{6}\,\alpha_2 \beta_2-2 \alpha_3 \beta_1 \\
     -2 \alpha_1 \beta_4-2 \alpha_4 \beta_1+\sqrt{6}\,\alpha_5 \beta_5 \\
     \alpha_1 \beta_5+\sqrt{6}\,\alpha_2 \beta_4+\sqrt{6}\,\alpha_4 \beta_2+\alpha_5 \beta_1
                     \end{pmatrix}\\[11mm]
\quad \mathbf{5}_2   & \begin{pmatrix}
     2 \alpha_1 \beta_1-2 \alpha_2 \beta_5+\alpha_3 \beta_4+\alpha_4\beta_3-2 \alpha_5 \beta_2 \\
     -2 \alpha_1 \beta_2-2 \alpha_2 \beta_1+\sqrt{6}\,\alpha_4 \beta_4 \\
     \alpha_1 \beta_3+\alpha_3 \beta_1+\sqrt{6}\,\alpha_4 \beta_5+\sqrt{6}\,\alpha_5 \beta_4 \\
     \alpha_1 \beta_4+\sqrt{6}\,\alpha_2 \beta_3+\sqrt{6}\,\alpha_3 \beta_2+\alpha_4 \beta_1 \\
     -2 \alpha_1 \beta_5+\sqrt{6}\,\alpha_3 \beta_3-2 \alpha_5 \beta_1
                     \end{pmatrix}
\end{array}\right.
\end{array}
\end{align}

%%%%%%%%%%%%%%%%%%%%%%%
\section{Modular Form Multiplets}
\label{app:multiplets}
%%%%%%%%%%%%%%%%%%%%%%%
%=======================
\subsection{Lowest Weight Multiplets}
\label{app:lw_multiplets}
%=======================

For the groups $\Gamma_N$ with $N \leq 5$, the lowest (non-trivial) weight
modular multiplets can be constructed from linear combinations $Y(a_i|\tau)$ of logarithmic derivatives of some seed
functions. These functions are the Dedekind eta $\eta(\tau)$ and Jacobi theta $\theta_3(z(\tau),t(\tau))$ functions,%
\footnote{For the properties of this last special function, see,~\mbox{e.g.~\cite{Farkas:2001th,Kharchev:2015tv}}.
In the notations of \mbox{Ref.~\cite{Farkas:2001th}}, $\theta_3 \equiv \theta\begin{bmatrix}0\\0\end{bmatrix}$.}
with modified arguments.
The Dedekind eta is defined as an infinite product,
\begin{align}
\eta(\tau) = q^{1/24} \,\prod_{n=1}^\infty \,\left(1-q^n\right)\,,\qquad  q = e^{2\pi i \tau}\,,
\end{align}
while the Jacobi $\theta_3$ can be written as the series
\begin{align}
\theta_3(z,\tau) = \sum_{n=-\infty}^\infty \,\tilde q^{\,n^2} \,e^{2 \pi i n z} \,,\qquad  \tilde q = e^{\pi i \tau}\,.
\end{align}

Below, for each $N \leq 5$, we reproduce explicit expressions for the lowest weight modular multiplets $Y^{(N,k=2)}_\mathbf{r}$
as vectors of the aforementioned log-derivatives.
Note that these multiplets are given in the symmetric basis of Appendix~\ref{app:sym_basis}.
For these groups, the dimensions of linear spaces at different weights 
and the irreducible representations arising at lowest weight
are summarised in Table~\ref{tab:low_irreps}.
\begin{table}
\centering
\renewcommand{\arraystretch}{1.2}
\begin{tabular}{lcccc}
  \toprule 
  $\Gamma_N$ & $\Gamma_2 \simeq S_3$ & $\Gamma_3 \simeq A_4$ & $\Gamma_4 \simeq S_4$ & $\Gamma_5 \simeq A_5$ \\
  \midrule
  Linear space dim.~at weight $k$ & $k/2 + 1$ & $k + 1$ & $2k + 1$ & $5k + 1$  \\
  Linear space dim.~at weight $2$ & $2$ & $3$ & $5$ & $11$ \\
  Lowest weight irreps  & $\mathbf{2}$ & $\mathbf{3}$ & $\mathbf{2}, \mathbf{3'}$ & $\mathbf{3}, \mathbf{3'}, \mathbf{5}$ \\
  \bottomrule
\end{tabular}
\caption{Dimensions of the linear spaces of modular forms at generic weight $k$ and at the lowest weight ($k=2$)~\cite{Gunning:1962lmf},
including the irrep breakdown at the lowest weight~\cite{Kobayashi:2018vbk,Feruglio:2017spp,Penedo:2018nmg,Novichkov:2018nkm},
for the finite modular groups $\Gamma_{N}$ with $N\leq 5$.}
\label{tab:low_irreps}
\end{table}

\subsubsection{\texorpdfstring{$\Gamma_2 \simeq S_3$}{Γ2=S3}}
The lowest weight multiplet for $\Gamma_2$ was derived in Ref.~\cite{Kobayashi:2018vbk}, and reads (up to normalisation):
\begin{align}
Y_\mathbf{2}^{(2,2)}(\tau) 
&=i \left(\begin{array}{c}
Y^{(2)}(1,1,-2|\tau)\\
Y^{(2)}(\sqrt{3},-\sqrt{3},0|\tau)
\end{array}\right)\,,
\end{align}
with
\begin{align}
Y^{(2)}(a_1,a_2,a_3|\tau) &= \frac{\di}{\di\tau} \bigg[
a_1 \log \eta\left(\frac{\tau}{2}\right) +
a_2 \log \eta\left(\frac{\tau+1}{2}\right) +
a_3 \log \eta\left(2\tau\right) 
\bigg]\,.
\end{align}

\subsubsection{\texorpdfstring{$\Gamma_3 \simeq A_4$}{Γ3=A4}}
The lowest weight multiplet for $\Gamma_3$ was derived in Ref.~\cite{Feruglio:2017spp}, and reads (up to normalisation):
\begin{align}
Y_\mathbf{3}^{(3,2)}(\tau) 
&= i \left(\begin{array}{c}
Y^{(3)}(1/2,1/2,1/2,-3/2|\tau)\\
-Y^{(3)}(1,\omega^2,\omega,0|\tau)\\
-Y^{(3)}(1,\omega,\omega^2,0|\tau)
\end{array}\right)\,,
\end{align}
with $\omega = e^{2\pi i/3}$ and
\begin{align}
Y^{(3)}(a_1,\dots,a_4|\tau) &= \frac{\di}{\di\tau} \bigg[
a_1 \log \eta\left(\frac{\tau}{3}\right) +
a_2 \log \eta\left(\frac{\tau+1}{3}\right) +
a_3 \log \eta\left(\frac{\tau+2}{3}\right) +
a_4 \log \eta\left(3\tau\right) 
\bigg]\,.
\end{align}

\subsubsection{\texorpdfstring{$\Gamma_4 \simeq S_4$}{Γ4=S4}}
\label{app:level4weight2}
The lowest weight multiplets for $\Gamma_4$ were derived in Ref.~\cite{Penedo:2018nmg}, in a non-symmetric basis for the representation of group generators. In the symmetric basis we here consider, they read (up to normalisation):
\begin{align}
Y_\mathbf{2}^{(4,2)}(\tau) 
&=i \left(\begin{array}{c}
Y^{(4)}(1,1,-1/2,-1/2,-1/2,-1/2|\tau)\\
Y^{(4)}(0,0,\sqrt{3}/2,-\sqrt{3}/2,\sqrt{3}/2,-\sqrt{3}/2|\tau)
\end{array}\right)\,,\\[2mm]
Y_\mathbf{3'}^{(4,2)}(\tau) 
&=i \left(\begin{array}{c}
Y^{(4)}(1,-1,0,0,0,0|\tau)\\
Y^{(4)}(0,0,-1/\sqrt{2},i/\sqrt{2},1/\sqrt{2},-i/\sqrt{2}|\tau)\\
Y^{(4)}(0,0,-1/\sqrt{2},-i/\sqrt{2},1/\sqrt{2},i/\sqrt{2}|\tau)
\end{array}\right)\,,
\end{align}
with
\begin{align}
Y^{(4)}(a_1,\dots,a_6|\tau) &= \frac{\di}{\di\tau} \bigg[
a_1 \log \eta\left(\tau+\frac{1}{2}\right) +
a_2 \log \eta\left(4\tau\right) +
a_3 \log \eta\left(\frac{\tau}{4}\right) \nonumber\\
&+ a_4 \log \eta\left(\frac{\tau+1}{4}\right) + 
a_5 \log \eta\left(\frac{\tau+2}{4}\right) +
a_6 \log \eta\left(\frac{\tau+3}{4}\right)\bigg]\,.
\end{align}

\subsubsection{\texorpdfstring{$\Gamma_5 \simeq A_5$}{Γ5=A5}}
The lowest weight multiplets for $\Gamma_5$ were derived in Ref.~\cite{Novichkov:2018nkm}, and read (up to normalisation):
\begin{align}
Y_\mathbf{5}^{(5,2)}(\tau) 
&=i\left(\begin{array}{c}
-\frac{1}{\sqrt{6}}Y^{(5)}\left(-5,1,1,1,1,1;-5,1,1,1,1,1\middle|\tau\right)\\
Y^{(5)}(0,1,\zeta^4,\zeta^3,\zeta^2,\zeta\,;\,0,1,\zeta^4,\zeta^3,\zeta^2,\zeta\,|\,\tau)\\
Y^{(5)}(0,1,\zeta^3,\zeta,\zeta^4,\zeta^2\,;\,0,1,\zeta^3,\zeta,\zeta^4,\zeta^2\,|\,\tau)\\
Y^{(5)}(0,1,\zeta^2,\zeta^4,\zeta,\zeta^3\,;\,0,1,\zeta^2,\zeta^4,\zeta,\zeta^3\,|\,\tau)\\
Y^{(5)}(0,1,\zeta,\zeta^2,\zeta^3,\zeta^4\,;\,0,1,\zeta,\zeta^2,\zeta^3,\zeta^4\,|\,\tau)
\end{array}\right)\,,\\[2mm]
Y_\mathbf{3}^{(5,2)}(\tau) 
&=i\left(\begin{array}{c}
\frac{1}{\sqrt{2}}Y^{(5)}\left(-\sqrt{5},-1,-1,-1,-1,-1;\sqrt{5},1,1,1,1,1\middle|\tau\right)\\
Y^{(5)}(0,1,\zeta^4,\zeta^3,\zeta^2,\zeta\,;\,0,-1,-\zeta^4,-\zeta^3,-\zeta^2,-\zeta\,|\,\tau)\\
Y^{(5)}(0,1,\zeta,\zeta^2,\zeta^3,\zeta^4\,;\,0,-1,-\zeta,-\zeta^2,-\zeta^3,-\zeta^4\,|\,\tau)
\end{array}\right)\,, \\[2mm]
Y_\mathbf{3'}^{(5,2)}(\tau) 
&=i\left(\begin{array}{c}
\frac{1}{\sqrt{2}}Y^{(5)}\left(\sqrt{5},-1,-1,-1,-1,-1;-\sqrt{5},1,1,1,1,1\middle|\tau\right)\\
Y^{(5)}(0,1,\zeta^3,\zeta,\zeta^4,\zeta^2\,;\,0,-1,-\zeta^3,-\zeta,-\zeta^4,-\zeta^2\,|\,\tau)\\
Y^{(5)}(0,1,\zeta^2,\zeta^4,\zeta,\zeta^3\,;\,0,-1,-\zeta^2,-\zeta^4,-\zeta,-\zeta^3\,|\,\tau)
\end{array}\right)\,,
\end{align}
with $\zeta = e^{2\pi i/5}$ and
\begin{align}
Y^{(5)}(c_{1,-1},\ldots,c_{1,4};c_{2,-1},\ldots,c_{2,4}|\tau) = \sum_{i,j} c_{i,j}\,
\frac{\di}{\di\tau}\log\alpha_{i,j}(\tau)\,,
\end{align}
the corresponding seed functions being
\begin{equation}
\begin{aligned}[c]
\alpha_{1,-1}(\tau) \,&=\, \theta_3\left( \frac{\tau+1}{2}, 5\tau\right)
\,, \\
\alpha_{1,0}(\tau) \,&=\, \theta_3\left( \frac{\tau+9}{10}, \frac{\tau}{5}\right)
\,, \\
\alpha_{1,1}(\tau) \,&=\, \theta_3\left( \frac{\tau}{10}, \frac{\tau+1}{5}\right)
\,, \\
\alpha_{1,2}(\tau) \,&=\, \theta_3\left( \frac{\tau+1}{10}, \frac{\tau+2}{5}\right)
\,, \\
\alpha_{1,3}(\tau) \,&=\, \theta_3\left( \frac{\tau+2}{10}, \frac{\tau+3}{5}\right)
\,, \\
\alpha_{1,4}(\tau) \,&=\, \theta_3\left( \frac{\tau+3}{10}, \frac{\tau+4}{5}\right)
\,,
\end{aligned}
\qquad
\begin{aligned}[c]
\alpha_{2,-1}(\tau) \,&=\, e^{2 \pi i \tau / 5} \, 
\theta_3\left( \frac{3\tau+1}{2}, 5\tau\right)
\,, \\
\alpha_{2,0}(\tau) \,&=\, \theta_3\left( \frac{\tau+7}{10}, \frac{\tau}{5}\right)
\,, \\
\alpha_{2,1}(\tau) \,&=\, \theta_3\left( \frac{\tau+8}{10}, \frac{\tau+1}{5}\right)
\,, \\
\alpha_{2,2}(\tau) \,&=\, \theta_3\left( \frac{\tau+9}{10}, \frac{\tau+2}{5}\right)
\,, \\
\alpha_{2,3}(\tau) \,&=\, \theta_3\left( \frac{\tau}{10}, \frac{\tau+3}{5}\right)
\,, \\
\alpha_{2,4}(\tau) \,&=\, \theta_3\left( \frac{\tau+1}{10}, \frac{\tau+4}{5}\right)
\,.
\end{aligned}
\end{equation}

%=======================
\subsection{Bases for Spaces of Lowest Weight Forms and Their \texorpdfstring{$q$}{q}-Expansions}
\label{app:qexpansions}
%=======================
For each $\Gamma_N$, one can obtain $q$-expansions for a basis $b_i^{(N)}$ (the so-called Miller-like basis)
of the space of lowest weight modular forms from the SageMath algebra system~\cite{SageMath:2018} (see, e.g.,~\cite{Novichkov:2018nkm}).
The dimensions of these linear spaces have been given in Table~\ref{tab:low_irreps}.
Below,  for each $N\leq 5$, we present these expansions
as well as the decompositions of the lowest weight multiplets of Appendix~\ref{app:lw_multiplets} (given in the symmetric basis of Appendix~\ref{app:sym_basis})
in terms of the basis vectors $b_i^{(N)} \equiv b_i^{(N)}(\tau)$.

\subsubsection{\texorpdfstring{$\Gamma_2 \simeq S_3$}{Γ2=S3}}
For $\Gamma_2$, the Miller-like basis reads:
\begin{equation}
\begin{aligned}
b_1^{(2)} &= 1 + 24q_2^2 + 24q_2^4 + 96q_2^6 + 24q_2^8 + 144q_2^{10} + 96q_2^{12} + 192q_2^{14} + 24q_2^{16} + 312q_2^{18} + 144q_2^{20}+\ldots\,,\\
b_2^{(2)} &= q_2 + 4q_2^3 + 6q_2^5 + 8q_2^7 + 13q_2^9 + 12q_2^{11} + 14q_2^{13} + 24q_2^{15} + 18q_2^{17} + 20q_2^{19}+\ldots\,,
\end{aligned}
\end{equation}
with $q_2 \equiv e^{2 \pi i\,\tau/2} = e^{\pi i\,\tau}$.
The lowest weight modular multiplet of $\Gamma_2$ can be written in terms of the above vectors, namely (see also~\cite{Kobayashi:2018vbk}):
\begin{align}
Y^{(2,2)}_\mathbf{2} (\tau) &=
2 \pi\,
\left(\begin{array}{c}
b_1^{(2)}(\tau) / 8\\[1mm]
\sqrt{3}\,b_2^{(2)}(\tau)
\end{array}\right)\,.
\end{align}
In the present Appendix, modular multiplets are
properly normalised, guaranteeing $Y(-\tau^*) = Y^*(\tau)$.

\subsubsection{\texorpdfstring{$\Gamma_3 \simeq A_4$}{Γ3=A4}}
For $\Gamma_3$, the Miller-like basis reads:
\begin{equation}
\begin{aligned}
b_1^{(3)} &= 1 + 12q_3^3 + 36q_3^6 + 12q_3^9 + 84q_3^{12} + 72q_3^{15} + 36q_3^{18} + 96q_3^{21} + 180q_3^{24} + 12q_3^{27} + 216q_3^{30} +\ldots\,,\\
b_2^{(3)} &= q_3 + 7q_3^4 + 8q_3^7 + 18q_3^{10} + 14q_3^{13} + 31q_3^{16} + 20q_3^{19} + 36q_3^{22} + 31q_3^{25} + 56q_3^{28} +\ldots\,,\\
b_3^{(3)} &= q_3^2 + 2q_3^5 + 5q_3^8 + 4q_3^{11} + 8q_3^{14} + 6q_3^{17} + 14q_3^{20} + 8q_3^{23} + 14q_3^{26} + 10q_3^{29} +\ldots\,,
\end{aligned}
\end{equation}
with $q_3 \equiv e^{2 \pi i\,\tau/3}$.
The lowest weight modular multiplet of $\Gamma_3$ can be written in terms of the above vectors, namely (see also~\cite{Feruglio:2017spp}):
\begin{align}
Y^{(3,2)}_\mathbf{3} (\tau) &=
\frac{\pi}{3} \,
\left(\begin{array}{c}
b_1^{(3)}(\tau)\\[1mm]
-6\,b_2^{(3)}(\tau)\\[1mm]
-18\,b_3^{(3)}(\tau)
\end{array}\right)\,.
\end{align}

\subsubsection{\texorpdfstring{$\Gamma_4 \simeq S_4$}{Γ4=S4}}
For $\Gamma_4$, the Miller-like basis reads:
\begin{equation}
\begin{aligned}
b_1^{(4)} &= 1 + 24 q_4^8 + 24 q_4^{16} + 96 q_4^{24} + 24q_4^{32} + 144q_4^{40} + \ldots\,,\\
b_2^{(4)} &= q_4 + 6q_4^5 + 13 q_4^9 + 14 q_4^{13} + 18q_4^{17} + 32q_4^{21} + 31q_4^{25} + 30q_4^{29} + 48q_4^{33} + 38q_4^{37} + \ldots\,,\\
b_3^{(4)} &= q_4^2 + 4q_4^6 + 6q_4^{10} + 8q_4^{14} + 13q_4^{18} + 12q_4^{22} + 14q_4^{26} + 24q_4^{30} + 18q_4^{34} + 20q_4^{38} + \ldots\,,\\
b_4^{(4)} &= q_4^3 + 2q_4^7 + 3q_4^{11} + 6q_4^{15} + 5q_4^{19} + 6q_4^{23} + 10q_4^{27} + 8q_4^{31} + 12q_4^{35} + 14q_4^{39} + \ldots\,,\\
b_5^{(4)} &= q_4^4 + 4q_4^{12} + 6q_4^{20} + 8q_4^{28} + 13q_4^{36} + \ldots\,,
\end{aligned}
\end{equation}
with $q_4 \equiv e^{2 \pi i\,\tau/4} = e^{\pi i\,\tau/2}$.
The lowest weight modular multiplets of $\Gamma_4$ can be written in terms of the above vectors, namely:
\begin{align}
Y^{(4,2)}_\mathbf{2} (\tau) &=
-3 \pi \,
\left(\begin{array}{c}
b_1^{(4)}(\tau)/8+3\,b_5^{(4)}(\tau)\\[1mm]
-\sqrt{3}\,b_3^{(4)}(\tau)
\end{array}\right)\,,\\[2mm]
Y^{(4,2)}_\mathbf{3'} (\tau) &=
-\pi\,
\left(\begin{array}{c}
-b_1^{(4)}(\tau)/4+2\,b_5^{(4)}(\tau)\\[1mm]
\sqrt{2}\,b_2^{(4)}(\tau)\\[1mm]
4\sqrt{2}\,b_4^{(4)}(\tau)
\end{array}\right)\,.
\end{align}

\subsubsection{\texorpdfstring{$\Gamma_5 \simeq A_5$}{Γ5=A5}}
For $\Gamma_5$, the Miller-like basis reads:
\begin{equation}
\begin{aligned}
b_1^{(5)} &= 1 + 60 q_5^{15} - 120 q_5^{20} + 240 q_5^{25} - 300 q_5^{30} + 300 q_5^{35} - 180 q_5^{45} + 240 q_5^{50}+\ldots\,,\\
b_2^{(5)} &=  q_5 + 12 q_5^{11} + 7 q_5^{16} + 8 q_5^{21} + 6 q_5^{26} + 32 q_5^{31} + 7 q_5^{36} + 42 q_5^{41} + 12 q_5^{46}+\ldots\,, \\
b_3^{(5)} &=  q_5^{2} + 12 q_5^{12} - 2 q_5^{17} + 12 q_5^{22} + 8 q_5^{27} + 21 q_5^{32} - 6 q_5^{37} + 48 q_5^{42} - 8 q_5^{47}+\ldots\,, \\
b_4^{(5)} &=  q_5^{3} + 11 q_5^{13} - 9 q_5^{18} + 21 q_5^{23} - q_5^{28} + 12 q_5^{33} + 41 q_5^{43} - 29 q_5^{48}+\ldots\,, \\
b_5^{(5)} &=  q_5^{4} + 9 q_5^{14} - 12 q_5^{19} + 29 q_5^{24} - 18 q_5^{29} + 17 q_5^{34} + 8 q_5^{39} + 12 q_5^{44} - 16 q_5^{49}+\ldots\,, \\
b_6^{(5)} &=  q_5^{5} + 6 q_5^{15} - 9 q_5^{20} + 27 q_5^{25} - 28 q_5^{30} + 30 q_5^{35} - 11 q_5^{45} + 26 q_5^{50}+\ldots\,,\\ 
b_7^{(5)} &=  q_5^{6} + 2 q_5^{16} + 2 q_5^{21} + 3 q_5^{26} + 7 q_5^{36} + 5 q_5^{46}+\ldots\,, \\
b_8^{(5)} &=  q_5^{7} - q_5^{12} + 3 q_5^{17} + 2 q_5^{27} + 7 q_5^{37} - 6 q_5^{42} + 9 q_5^{47}+\ldots\,,\\ 
b_9^{(5)} &=  q_5^{8} - 2 q_5^{13} + 5 q_5^{18} - 4 q_5^{23} + 4 q_5^{28} + 4 q_5^{38} - 8 q_5^{43} + 16 q_5^{48}+\ldots\,,\\ 
b_{10}^{(5)} &=  q_5^{9} - 3 q_5^{14} + 8 q_5^{19} - 11 q_5^{24} + 12 q_5^{29} - 5 q_5^{34} + 13 q_5^{49}+\ldots\,, \\
b_{11}^{(5)} &=  q_5^{10} - 4 q_5^{15} + 12 q_5^{20} - 22 q_5^{25} + 30 q_5^{30} - 24 q_5^{35} + 5 q_5^{40} + 18 q_5^{45} - 21 q_5^{50}+\ldots\,,
\end{aligned}
\end{equation}
with $q_5 \equiv e^{2 \pi i\,\tau/5}$.
The lowest weight modular multiplets of $\Gamma_5$ can be written in terms of the above vectors, namely (see also~\cite{Novichkov:2018nkm}):
\begin{align}
Y^{(5,2)}_\mathbf{3} (\tau) &=
-2 \sqrt{5} \pi \,
\left(\begin{array}{c}
\left(b_1^{(5)}(\tau) - 30\, b_6^{(5)}(\tau) - 20\, b_{11}^{(5)}(\tau) \right) / (5\sqrt{2})\\[1mm]
 b_2^{(5)}(\tau) + 2\, b_7^{(5)}(\tau)\\[1mm]
 3\,b_5^{(5)}(\tau) + 7\, b_{10}^{(5)}(\tau)
\end{array}\right)\,,\\[2mm]
Y^{(5,2)}_\mathbf{3'} (\tau) &=
-2 \sqrt{5} \pi \,
\left(\begin{array}{c}
-\left(b_1^{(5)}(\tau) + 20\,b_6^{(5)}(\tau) + 30\,b_{11}^{(5)}(\tau) \right) / (5\sqrt{2})\\[1mm]
 b_3^{(5)}(\tau) + 6\,b_8^{(5)}(\tau)\\[1mm]
 2\,b_4^{(5)}(\tau) + 5\,b_9^{(5)}(\tau)
\end{array}\right)\,,\\[2mm]
Y^{(5,2)}_\mathbf{5} (\tau) &=
2 \pi \,
\left(\begin{array}{c}
-\left(b_1^{(5)}(\tau) + 6\,b_6^{(5)}(\tau) + 18\,b_{11}^{(5)}(\tau)\right)/\sqrt{6}\\[1mm]
b_2^{(5)}(\tau) + 12\, b_7^{(5)}(\tau)\\[1mm]
3\, b_3^{(5)}(\tau) + 8\, b_8^{(5)}(\tau)\\[1mm]
4\, b_4^{(5)}(\tau) + 15\, b_9^{(5)}(\tau) \\[1mm]
7 \, b_5^{(5)}(\tau) + 13\, b_{10}^{(5)}(\tau)
\end{array}\right)\,.
\end{align}

%=======================
\subsection{Higher Weight Multiplets for \texorpdfstring{$\Gamma_4 \simeq S_4$}{Γ4=S4} in the Symmetric Basis}
\label{app:higher_weights}
%=======================
Multiplets of higher weight $Y^{(N,k>2)}_{\mathbf{r},\mu}$, with the index $\mu$ labelling linearly independent multiplets,
may be obtained from those of lower weight via tensor products.
For $\Gamma_3$, multiplets of weight up to 6 are found in Ref.~\cite{Feruglio:2017spp},
while for $\Gamma_4$ and $\Gamma_5$, multiplets of weight up to 10 are given in Refs.~\cite{Novichkov:2018ovf} and~\cite{Novichkov:2018nkm}, respectively.

Since in~\cite{Novichkov:2018ovf} the considered basis for $\Gamma_4$ is not a symmetric one, we here
present explicit expressions for higher weight multiplets in the symmetric basis of Appendix~\ref{app:sym_basis}.
Making use of the shorthands $Y^{(4,2)}_\mathbf{2} = (Y_1, Y_2)^T$ and $Y^{(4,2)}_\mathbf{3'} = (Y_3, Y_4, Y_5)^T$,
one has at weight 4,
\begin{equation}
\begin{aligned}
Y^{(4,4)}_\mathbf{1} = Y_1^2 + Y_2^2\,,&
\qquad
Y^{(4,4)}_\mathbf{2} = \begin{pmatrix}
   Y_2^2-Y_1^2 \\
2\,Y_1 Y_2
\end{pmatrix}, \\
Y^{(4,4)}_\mathbf{3} = \begin{pmatrix}
-2\,Y_2 Y_3 \\
 \sqrt{3} \, Y_1 Y_5 + Y_2 Y_4 \\
 \sqrt{3} \, Y_1 Y_4 + Y_2 Y_5
\end{pmatrix},&
\qquad
Y^{(4,4)}_\mathbf{3'} = \begin{pmatrix}
 2\,Y_1 Y_3 \\
 \sqrt{3} Y_2 Y_5 - Y_1 Y_4 \\
 \sqrt{3} Y_2 Y_4 - Y_1 Y_5 
\end{pmatrix},
\end{aligned}
\end{equation}
at weight 6,
\begin{equation}
\begin{aligned}
Y^{(4,6)}_\mathbf{1} = Y_1 \left( 3\, Y_2^2-Y_1^2\right) \,, &
\qquad
Y^{(4,6)}_\mathbf{1'} = Y_2 \left( 3\, Y_1^2-Y_2^2\right) \,, \\[1mm]
Y^{(4,6)}_\mathbf{2} = 
\left(Y_1^2+Y_2^2\right) \begin{pmatrix}
 Y_1  \\
 Y_2 
\end{pmatrix}, &
\qquad
Y^{(4,6)}_\mathbf{3} = \begin{pmatrix}
  Y_1 \left(Y_4^2-Y_5^2\right) \\
  Y_3 \left(Y_1 Y_5+ \sqrt{3}\, Y_2 Y_4 \right) \\
 -Y_3 \left(Y_1 Y_4 + \sqrt{3}\, Y_2 Y_5\right)
\end{pmatrix}, \\
Y^{(4,6)}_{\mathbf{3'},1} =
\left(Y_1^2+Y_2^2\right)  \begin{pmatrix}
 Y_3 \\
 Y_4 \\
 Y_5 
\end{pmatrix},&
\qquad
Y^{(4,6)}_{\mathbf{3'},2} = \begin{pmatrix}
  Y_2 \left(Y_5^2-Y_4^2\right) \\
 -Y_3 \left(Y_2 Y_5 - \sqrt{3}\, Y_1 Y_4\right) \\
  Y_3 \left(Y_2 Y_4 - \sqrt{3}\, Y_1 Y_5\right)
\end{pmatrix},
\end{aligned}
\end{equation}
at weight 8,
\begin{equation}
\begin{aligned}
Y^{(4,8)}_\mathbf{1} = \left(Y_1^2 + Y_2^2\right)^2 \,, & \\[1mm]
Y^{(4,8)}_{\mathbf{2},1} =
\left(Y_1^2+Y_2^2\right) \begin{pmatrix}
 Y_2^2- Y_1^2  \\
 2\,Y_1 Y_2 
\end{pmatrix}, &
\qquad
Y^{(4,8)}_{\mathbf{2},2} = 
\left(Y_1^2 - 3\, Y_2^2\right) \begin{pmatrix}
 Y_1^2 \\
 Y_1 Y_2 
\end{pmatrix}, \\
Y^{(4,8)}_{\mathbf{3},1} =
\left(Y_1^2+Y_2^2\right)  \begin{pmatrix}
 -2\, Y_2 Y_3 \\
 Y_2 Y_4 + \sqrt{3}\, Y_1 Y_5 \\
 Y_2 Y_5 + \sqrt{3}\, Y_1 Y_4
\end{pmatrix},&
\qquad
Y^{(4,8)}_{\mathbf{3},2} = 
Y_2 \left(Y_2^2 - 3\, Y_1^2\right) \begin{pmatrix}
 Y_3 \\
 Y_4 \\
 Y_5 
\end{pmatrix}, \\
Y^{(4,8)}_{\mathbf{3'},1} =
\left(Y_1^2+Y_2^2\right) \begin{pmatrix}
 -2\, Y_1 Y_3 \\
 Y_1 Y_4 - \sqrt{3}\, Y_2 Y_5\\
 Y_1 Y_5 - \sqrt{3}\, Y_2 Y_4
\end{pmatrix},&
\qquad
Y^{(4,8)}_{\mathbf{3'},2} = 
Y_1\left(Y_1^2 - 3\, Y_2^2\right) \begin{pmatrix}
 Y_3 \\
 Y_4 \\
 Y_5 
\end{pmatrix},
\end{aligned}
\end{equation}
and at weight 10,
\begin{equation}
\begin{aligned}
Y^{(4,10)}_\mathbf{1} = Y_1 \left(Y_1^2-3\, Y_2^2\right) \left(Y_1^2+Y_2^2\right) \,, & 
\qquad
Y^{(4,10)}_\mathbf{1'} =  Y_2 \left(Y_2^2-3\, Y_1^2\right) \left(Y_1^2+Y_2^2\right)  \,, \\[1mm]
Y^{(4,10)}_{\mathbf{2},1} =
\left(Y_1^2+Y_2^2\right)^2 \begin{pmatrix}
 Y_1  \\
 Y_2 
\end{pmatrix}, &
\qquad
Y^{(4,10)}_{\mathbf{2},2} = 
\left(Y_1^2 - 3\, Y_2^2\right) \begin{pmatrix}
 Y_1^3-Y_1 Y_2^2 \\
 -2\, Y_1^2 Y_2 
\end{pmatrix}, \\
Y^{(4,10)}_{\mathbf{3},1} =
Y_1 \left(Y_1^2 - 3\, Y_2^2\right) \begin{pmatrix}
 -2\, Y_2 Y_3 \\
 Y_2 Y_4 + \sqrt{3}\, Y_1 Y_5 \\
 Y_2 Y_5 + \sqrt{3}\, Y_1 Y_4
\end{pmatrix},&
\qquad
Y^{(4,10)}_{\mathbf{3},2} = 
Y_2 \left(Y_2^2 - 3\, Y_1^2\right) \begin{pmatrix}
 -2\, Y_1 Y_3 \\
 Y_1 Y_4 - \sqrt{3}\, Y_2 Y_5\\
 Y_1 Y_5 - \sqrt{3}\, Y_2 Y_4
\end{pmatrix}, \\
Y^{(4,10)}_{\mathbf{3'},1} =
Y_1 \left(Y_1^2 - 3\, Y_2^2\right) \begin{pmatrix}
 -2\, Y_1 Y_3 \\
 Y_1 Y_4 - \sqrt{3}\, Y_2 Y_5\\
 Y_1 Y_5 - \sqrt{3}\, Y_2 Y_4
\end{pmatrix},&
\qquad
Y^{(4,10)}_{\mathbf{3'},2} = 
Y_2 \left(Y_2^2 - 3\, Y_1^2\right)  \begin{pmatrix}
 -2\, Y_2 Y_3 \\
 Y_2 Y_4 + \sqrt{3}\, Y_1 Y_5 \\
 Y_2 Y_5 + \sqrt{3}\, Y_1 Y_4
\end{pmatrix}, \\
Y^{(4,10)}_{\mathbf{3'},3} =
\left(Y_1^2+Y_2^2\right)^2 \begin{pmatrix}
 Y_3 \\
 Y_4 \\
 Y_5 
\end{pmatrix}.
\end{aligned}
\end{equation}

%%%%%%%%%%%%%%%%%%%%%%%
\bibliographystyle{JHEPwithnote}
\bibliography{bibliography}
%%%%%%%%%%%%%%%%%%%%%%%

%%%%%%%%%%%%%%%%%%%%%%%
\end{document}